\documentclass[prb,twocolumn,showpacs,preprintnumbers,amsmath,amssymb]
{revtex4}
\usepackage{graphicx}
\usepackage{dcolumn}
\usepackage{bm}

\begin{document}

\newcommand{\J}[1]{{\bf J:} #1 {\bf END}}
\newcommand{\AR}[1]{{\bf A:} #1 {\bf END}}
\newcommand{\w}{\omega}
\newcommand{\gz}{{^z\!\!\,g}}
\newcommand{\W}{\Omega}
\newcommand{\e}{\varepsilon}
\newcommand{\s}{\sigma}
\newcommand{\al}{\gamma}
\newcommand{\RE}{\text{Re}}
\newcommand{\IM}{\text{Im}}
\newcommand{\g}{\alpha}
\newcommand{\G}{\Gamma}
\newcommand{\Gt}{\Gamma_{2}}
\newcommand{\Gs}{\Gamma_{\!s}}
\newcommand{\Gl}{\Gamma_{1}}
\newcommand{\Gv}{\Gamma_{v}}
\newcommand{\IR}{E_{\rm{IR}}}
\newcommand{\TK}{T_{K}}
\newcommand{\I}{{\cal I}}
\newcommand{\K}{{\cal K}}
\newcommand{\Nf}{N(0)}
\newcommand{\la}{\lambda}
\newcommand{\La}{\Lambda}
\newcommand{\ind}[5]{{_{#1}}\!\!\!\!{^{#2}}#3{_{#4}}\!\!\!\!\!{^{#5}}}
\newcommand{\ovrl}{\overline}
\newcommand{\undl}{\underline}
\newcommand{\p}{\prime}
\newcommand{\up}{\uparrow}
\newcommand{\down}{\downarrow}
\newcommand{\gd}{g_{d}}
\newcommand{\glr}{g_{LR}}
\newcommand{\gll}{g_{LL}}
\newcommand{\grr}{g_{RR}}
\newcommand{\Gpf}{{\hat {\cal G}}}
\newcommand{\Lpf}{{\hat L}}
\newcommand{\GR}{{\cal G}^{R}}
\newcommand{\GA}{{\cal G}^{A}}
\newcommand{\GK}{{\cal G}^{K}}
\newcommand{\GL}{{\cal G}^{<}}
\newcommand{\GG}{{\cal G}^{>}}


\title{Nonequilibrium Transport through a Kondo Dot:\\
  Decoherence Effects.}
     
\author{J. Paaske$^1$, A. Rosch$^{1,2}$, J. Kroha$^3$, and P.
  W\"olfle$^1$}
     
\affiliation{$^1$\mbox{Institut f\"ur Theorie der Kondensierten
    Materie, Universit\"at Karlsruhe, 76128 Karlsruhe, Germany}\\
  $^2$Institut f\"ur Theoretische Physik, Universit\"at zu K\"oln,
  50937 K\"oln, Germany\\
  $^3$Physikalisches Institut, Universit\"at Bonn, 53115 Bonn,
  Germany}

\date{\today}

\begin{abstract}
  
  We investigate the effects of voltage induced spin-relaxation in a
  quantum dot in the Kondo regime. Using nonequilibrium perturbation
  theory, we determine the joint effect of self-energy and vertex
  corrections to the conduction electron T-matrix in the limit of
  transport voltage much larger than temperature. The logarithmic
  divergences, developing near the different chemical potentials of
  the leads, are found to be cut off by spin-relaxation rates,
  implying that the nonequilibrium Kondo-problem remains at weak
  coupling as long as voltage is much larger than the Kondo
  temperature.

\end{abstract}


\maketitle


Electron transport through quantum dots or point contacts possessing a
degenerate ground state (e.g. a  spin) is strongly influenced
by the Kondo effect~\cite{Hewson93}, provided the dot is in the
Coulomb blockade regime. In the linear response regime, the Kondo
resonance formed at the dot at sufficiently low temperature, i.e. at
or below the Kondo temperature $\TK$, allows for resonant tunneling,
thus removing the Coulomb blockade and leading to conductances near
the unitarity limit. This has been observed in various experiments on
quantum dot devices~\cite{Experiments}.
  
The Kondo resonance is quenched by either large temperature,
$T\gg\TK$, large magnetic field, $B\gg\TK$ or a large bias voltage,
$V\gg\TK$. However, the mechanism of how and why the Kondo effect is
suppressed is qualitatively different in the three cases. The Kondo
effect arises from resonant spin-flip scattering at the Fermi energy.
Temperature destroys the resonance mainly by smearing out the Fermi
surface, whereas a magnetic field lifts the degeneracy of the levels
on the dot and thereby prohibits resonant scattering. The effect of a
bias voltage, $V$, is more subtle. It induces a splitting of the Fermi
energies of the left, and the right lead.  However, this splitting
affects directly only resonant electron scattering from the left to
the right lead, but {\em not} any scattering which begins and ends on
the {\em same} lead. Yet these remaining resonant processes are
suppressed by a different effect: the voltage induces a current which
leads to noise and therefore to decoherence of resonant spin-flips. It
is the goal of this paper to study those decoherence effects in
detail.

In perturbation theory, the signature of Kondo physics are logarithmic
divergences arising from (principle value) integrals of the type
\begin{equation}
\int_{-D}^D\!d\w\, \frac{f(\w)}{\w}\sim\ln\frac{D}{\IR}
\end{equation}
where $f(\w)$ is the Fermi function, $D$ a high energy cutoff (i.e.
bandwidth) and $\IR$ some infrared cutoff. There are three rather
different ways to cut off the logarithm, and to destroy the Kondo
effect, corresponding to the three mechanisms discussed above. First,
temperature broadens $f(\w)$ leading to $\IR\sim T$. Second, a
magnetic field $B$ shifts the pole with respect to the Fermi-energy,
replacing $\frac{1}{\w}$ by $\frac{1}{\w-B}$, and in this case
$\IR\sim B$. The third way to quench the logarithm is to introduce a
finite decoherence rate $\Gs$, replacing $\frac{1}{\w}$ by
$\frac{\w}{\w^2+\Gs^2}$, implying $\IR\sim\Gs$.

The relaxation rate $\Gs=\Gs(V,B,T)$ and the associated decoherence
effects also exist in equilibrium. In the limit of vanishing bias
voltage and magnetic field, the scale $\Gs$ tends to a temperature
dependent (Korringa) rate\cite{Korringa50}, $\Gs(0,0,T)\ll T$, which
vanishes as $T\to 0$, allowing for the quantum coherent Kondo state to
be formed. In the case of a finite magnetic field and zero
temperature, a $B$- and spin-dependent rate\cite{Wang68},
$\G_{\!s,\sigma}(0,B,0)$ remains finite for the excited state
$\sigma=\downarrow$. In dynamic quantities it prohibits singular
behavior at $\w\sim B$ but it is not important for static quantities,
where $B$ eliminates all relevant singularities.  In the case of a
finite bias voltage $V$, however, the finite rate $\Gs(V,0,0)$ is
instrumental to cut off singularities even in {\em static} quantities
for $T,B \to 0$. The Kondo effect develops only to a certain extent,
depending on the ratio $V/T_K$.

Not only for a quantitative description of experiments in the regime
$V\gg\TK$, but even for a crude qualitative understanding of Kondo
physics out of equilibrium, it is necessary to identify the correct
relaxation rate $\Gs$. The question, how logarithmic contributions are
cut off, is essential to derive the correct perturbative
renormalization group description\cite{Schoeller00,Rosch03a} and to
identify regimes where novel strong-coupling physics is induced out of
equilibrium.

The importance of the broadening of the Zeeman levels was pointed out
three decades ago by Wolf and Losee\cite{Wolf69} in the context of the
Kondoesque tunneling anomaly observed in various tunnel junctions.
Incorporating a Korringa like, $T$ and $B$ dependent, spin-relaxation
rate into Appelbaum's perturbative formula for the
conductance\cite{Appelbaum66} was found to improve the agreement with
experiments considerably (cf. e.g.  Refs.~\onlinecite{Appelbaum72} and
\onlinecite{Bermon78}). Later, in the context of quantum dots, Meir
{\it et al.}\cite{Meir93} pointed out that, even at $T=B=0$, the
finite bias-voltage induces a broadening of the Zeeman levels. In
their self-consistent treatment of the Anderson model, using the
non-crossing approximation (NCA), this nonequilibrium broadening was
shown to suppress the Kondo peaks in the local density of states,
located at the two different Fermi levels.  In
Ref.~\onlinecite{Rosch01} we showed that this NCA relaxation rate is
sufficiently large to prohibit the flow towards strong coupling for
$V\gg T_K$. In a perturbative study of the effects of an ac-bias,
Kaminski {\it et al.}\cite{Kaminski99} argued that an irradiation
induced broadening serves to cut off the logarithmic divergence of the
conductance as $T$ and $V$ tend to zero.  Treatments of the Kondo
model\cite{Coleman01} and related problems\cite{Kiselev02} at large
voltages, which neglect the influence of decoherence, find strong
coupling effects even for $V\gg T_K$. Coleman {\it et
  al.}\cite{Coleman01} recently argued that this is the case because
$\Gs$ remains sufficiently small due to a (supposed) cancellation of
vertex and self-energy corrections.

To our knowledge, even to lowest order in perturbation theory, a
systematic calculation of the nonequilibrium decoherence rate is still
lacking. It is the objective of this paper to provide such a
calculation. This is a delicate matter since self-energy, and vertex
corrections may indeed cancel partially, and an infinite resummation
of perturbation theory is required. Recently\cite{Mao03, Shnirman03},
it was demonstrated that the Majorana fermion representation for the
local spin-$1/2$ circumvents this complication when calculating
spin-spin correlation functions. In this representation, such
correlators take the form of one-particle, rather than two-particle,
fermionic correlation functions, and consequently only self-energy
corrections have to be considered.  Whether this representation will
prove to be equally efficient for calculating other observables like
the conduction electron T-matrix or the conductance remains to be
seen.

Based on the conjecture that no unexpected cancellations occur, we
have recently developed a perturbative renormalization group
description\cite{Rosch03a} of the Kondo effect at large voltages. In
this approach, it was essential to include the effects of $\Gs$. For
usual quantum dots, the Kondo effect is sufficiently suppressed by
$\Gs$\cite{Rosch01, Rosch03a}, such that renormalized perturbation
theory remains applicable at all temperatures, provided $\ln(V/T_K)\gg
1$. We argued that $\Gs$, as a physically observable quantity, should
be identified with the transverse spin relaxation rate $\Gt=1/T_{2}$,
measuring the coherence property of the local spin (More precisely,
slightly different rates enter into various physical quantities, but
to leading order in $1/\ln[V/T_K]$ one can use $\Gs\approx\Gt$). In
this paper we show that within perturbation theory this is indeed the
case, thus confirming our initial conjecture. Note that in more
complex situations, for example in the case of coupled quantum dots,
$\Gs$ can be sufficiently small\cite{Rosch01} so that novel (strong
coupling) physics can be induced for large voltages.

In a preceding paper\cite{Paaske03a}, henceforth referred to as I, we
calculated perturbatively the local magnetization and the differential
conductance of a Kondo dot, including all leading logarithmic
corrections in the presence of finite $V$ and $B$. As effects of $\Gs$
are not included to this order, some logarithms were not cut off by
$V$ but appeared to diverge with $\ln(D/T)$ or $\ln(D/|V-B|)$. A
systematic calculation of the cut-off $\Gs$ requires a consistent
resummation of self-energy and vertex corrections. As will become
clear in the following, this is a formidable task, and we have
therefore concentrated on the quantity which appears to be most
tractable: the conduction electron T-matrix as a function of
frequency, in zero magnetic field.

In Sec.~\ref{model} we introduce the model and some conventions used
for the Keldysh perturbation theory. A combination of self-energy
corrections from Sec.~\ref{selfenergy} and vertex-corrections
calculated in Sec.~\ref{vertex} determines the spin-relaxation rate
(Sec.~\ref{susceptibility}). In Sec.~\ref{tmatrix} we show how this
decoherence rate cuts off logarithmic corrections in the T-matrix.  In
Sec.~\ref{anisotropic} we consider the case of anisotropic exchange
couplings and determine the exact combination of transverse and
longitudinal spin-relaxation rates which enters the logarithms in the
T-matrix. Appendices~\ref{app:crossed} and \ref{app:contract} contain
details pertaining to Sections~\ref{vertex} and \ref{tmatrix}.
Appendix~\ref{app:xray} investigates how power-law singularities of
the strongly anisotropic Kondo model are modified out of equilibrium
by mapping it to the nonequilibrium X-ray edge problem for vanishing
spin-flip coupling.

\section{Model and Method}\label{model}

We model the quantum dot by its local spin $\vec S$ $(S=\frac{1}{2})$,
coupled by the exchange interaction $J_{\g\g'}$ ($\g,\g'=L,R$) to
the conduction electrons in the left (L) and right (R) leads
\begin{eqnarray}
H&=&\sum_{\g,{\bf k},\s}(\e_{{\bf k}}-\mu_{\g})
c^{\dagger}_{\g{\bf k}\s}c_{\g{\bf k}\s}-g\mu_{B}B S_{z}\nonumber\\
& & + \!\sum_{\g,\g',{\bf k},{\bf k}',\s,\s'}\!\!\!\!\!J_{\g'\g}\,
\vec{S}\cdot\frac{1}{2}\,c^{\dagger}_{\g'{\bf k}'\s'}
\vec{\tau}_{\s'\s}c_{\g{\bf k}\s},\label{eq:hamilton}
\end{eqnarray}
where $J_{LR}$ describes a co-tunneling process transferring an
electron from the right to the left lead. Here $\mu_{L,R}=\pm eV/2$
are the chemical potentials of respectively the left and right leads,
$\vec \tau$ is the vector of Pauli matrices, $g\mu_BB$ the Zeeman
splitting of the local spin levels in a magnetic field $B$, and
$c_{\g{\bf k}\sigma}^{\dagger}$ creates an electron in lead $\g$ with
momentum ${\bf k}$ and spin $\sigma$. We will use dimensionless
coupling constants $g_{\g\g'} =N(0)J_{\g\g'}$, with $N(0)$ the density
of states per spin for the conduction electrons (assumed flat on the
scale $eV, g\mu_BB$).  We define for later use $\gd=(\gll+\grr)/2$,
$4g^{2}=\gll^{2}+\grr^{2}+2\glr^{2}$ and use units where
$\hbar=k_{B}=g\mu_{B}=e=1$.

In order to calculate observable quantities for the system with
Hamiltonian (\ref{eq:hamilton}), we find it convenient to use a
fermionic representation of the local spin operator,
\begin{equation}
\vec S =
\frac{1}{2} \sum_{\al\al{'}} f_\al^{\dagger}
\vec\tau_{\al\al{'}}f_{\al{'}},
\end{equation}
with canonical fermion creation and annihilation operators
$f_{\al}^{\dagger}$, $f_{\al}$, $\al=\up\down$, which allows a
conventional diagrammatic perturbation theory in the coupling constant
$g$. Since the physical Hilbert space must have singly occupied states
only, it is necessary to project out the empty and doubly occupied
local states. This is done by introducing a chemical potential $\la$
regulating the charge $Q=\sum_{\al}f_{\al}^{\dagger}f_{\al}$. Picking
out the contribution proportional to $e^{-\beta\lambda}$ and taking
the limit $\la\to\infty$, the constraint $Q=1$ can be enforced (for a
more detailed description of this method see I).

We will use the Keldysh Green function method for nonequilibrium
systems, following the notation of Ref.~\onlinecite{Rammer86}. Keldysh
matrix propagators are defined as
\begin{eqnarray}
\undl{G}=\left(\begin{array}{cc}
G^{R} & G^{K} \\
0     & G^{A} \\
\end{array}\right)\label{eq:matrix}
\end{eqnarray}
where $G^{R,A}$ and $G^K$ are the retarded, advanced and Keldysh
component Green functions, respectively. Spectral functions are found
as $A=i(G^{R}-G^{A})$, and the {\it greater} and {\it lesser}
functions as
\begin{equation}
G^{>/<}=(G^{K}\pm G^{R}\mp G^{A})/2.
\end{equation}

The local conduction electron ($ce$) Green functions at the dot in the
left and right leads, and the pseudo fermion ($pf$) Green function are
denoted by $G_{\g\s}^{ab}$ and ${\cal G}_{\al}^{cd}$, respectively,
with lead index $\g=L,R$, spin indices $\s,\al$, and Keldysh indices
$a,b,c,d$. A corresponding notation will be used for the self-energy
$\Sigma$, and its imaginary part, the self-energy broadening, is
denoted by $\G_\gamma=i(\Sigma^{R}_\gamma-\Sigma^{A}_\gamma)$.  The
interaction vertex has the following tensor structure in Keldysh space
\begin{equation}
\Lambda_{ab}^{cd}=\frac{1}{2}\left(\delta_{ab}\tau^{1}_{cd}
+\tau^{1}_{ab}\delta_{cd}\right),\label{eq:vertex0}
\end{equation}
where $a,b$ and $c,d$ refer to $pf$, and $ce$-lines, respectively.

Since we consider only nonequilibrium situations in a steady state,
time translation invariance holds, and the single-particle Green
functions depend only on one frequency. The bare $pf$ spectral
function is given by
\begin{equation}
{\cal A}_{\al}(\w)=2\pi\delta(\w+\al B/2),
\end{equation}
and the Keldysh component Green function is given as
\begin{equation}
\GK_{\al}(\w)=i{\cal A}_{\al}(\w)[2n_{\al\la}(\w)-1],
\end{equation}
where $n_{\al\la}(\w)$ denotes the $pf$ distribution function, given
by $n_{\al\la}(\w)=1/(e^{(\w+\la)/T}+1)$ in thermal equilibrium. We
shall also use the shorthand notation
\begin{equation}
M_{\al\la}=2n_{\al\la}(\w)-1.
\end{equation}

Assuming a constant conduction electron density of states $N(0)=1/2D$
and a bandwidth $2D$, the local $ce$ spectral function takes the form
\begin{equation}
A(\w)=2\pi N(0)\theta(D-|\w|)  
\end{equation}
with the step function $\theta(x)$.  The Keldysh component Green
function in lead $\g$ is then given by
\begin{equation}
G^{K}_{\g}(\w)=-iA(\w)\tanh\left(\frac{\w-\mu_{\g}}{2T}\right),  
\end{equation}
assuming the electrons in each lead to be in thermal equilibrium.

\section{Spin level broadening and spin relaxation rates}

The coupling of the local spin to the leads introduces a broadening of
the Zeeman levels, which depends on temperature, magnetic field and
bias voltage. In the pseudo fermion representation for the local spin,
the broadening is given by the imaginary part of the pseudo fermion
self-energy. This level broadening enters into the relaxation rates of
both the transverse spin components $(S_x,S_y)$, where it accounts for
the loss of phase coherence, and the longitudinal spin component
$(S_z)$, where it describes the relaxation of the local magnetization
following a change in the magnetic field. The observable spin
relaxation rates, $1/T_{2}$ and $1/T_{1}$, are defined through the
broadening of the resonance poles in the transverse, and longitudinal
dynamical spin susceptibilities, and their calculation requires vertex
corrections to be included in a consistent way.
 
Following a brief discussion of the $pf$ self-energy broadening, we
determine the renormalized $ce$-$pf$ interaction vertex in a
steady-state nonequilibrium situation. The resulting vertex functions
are used to calculate the transverse dynamical spin susceptibility,
and later, in Sec.~\ref{tmatrix}, they will serve as building blocks
for a calculation of the conduction electron T-matrix.

\subsection{Pseudo Fermion Decay Rates}\label{selfenergy}

In paper I (Ref.~\onlinecite{Paaske03a}), we determined the {\it
  on-shell} imaginary part of the pseudo fermion self-energy,
including leading logarithmic corrections.  For the purpose of this
paper, we will only need the second order rates, disregarding
logarithmic corrections. For $T=0$ one finds for $0\leq V<B$:
\begin{eqnarray}
\G_{\up}&=&\frac{\pi}{4}\glr^{2} V, \label{eq:rate1}\\
\G_{\down}&=&\G_{\up}+2\pi g^2 B, 
\end{eqnarray}
with $4g^{2}=\gll^{2}+\grr^{2}+2\glr^{2}$, whereas for $V>B\geq 0$:
\begin{eqnarray}
\G_{\up}&=&\frac{\pi}{4}\glr^{2} (3V-2B),\\
\G_{\down}&=&\G_{\up}+2\pi g^2 B.\label{eq:rate4}
\end{eqnarray} 
Notice that in the presence of a finite magnetic field, only the upper
spin-level, here corresponding to spin down, is broadened when $V=0$,
as one would expect from simple phase-space considerations.
Broadening of the lower spin-level (spin up) is due to virtual
transitions to the upper spin-level and occurs only in higher orders
in $g$.

For comparison, we list also the thermal decay rate for $V=B=0$:
\begin{equation}
\G_{\up,\down}=3\pi T g^{2}.
\end{equation}

\subsection{Vertex Corrections}\label{vertex}

Early work~\cite{Walker68,Woelfle71,Langreth72} on the dynamical
magnetic susceptibility of a single spin 1/2, demonstrated how
self-energy, and vertex corrections combine to yield the transverse,
and longitudinal relaxation rates $1/T_{2}$ and $1/T_{1}$. In
Ref.~\onlinecite{Walker68}, the vertex corrections were determined in
the approximation where the imaginary part of the $pf$ self-energy,
$\G$, is much smaller than temperature. A similar approach is possible
out of equilibrium, where it is the finite voltage, rather than
temperature, which determines the abundance of (inter-lead) conduction
electron particle-hole excitations. In the approximation where $\G\ll
V$, the dominant corrections to the $ce$-$pf$ interaction vertex
(Keldysh) tensor simplify substantially and the corresponding vertex
matrix equation can be solved analytically. Since we assume that
$V\gg\TK$, perturbation theory is valid and $\G\sim g^{2}V\ll V$ is
indeed a sound approximation. We shall consider the case where $T\ll
V$, which will best reveal the salient nonequilibrium features of the
problem. 

To calculate vertex corrections and their interplay with self-energy
diagrams, we have to solve the vertex equation depicted in
Fig.~\ref{fig:vertren}. We remind the reader that physical quantities
are proportional to $e^{-\beta \lambda}$ within our projection scheme.
Therefore we have to keep track of two contributions to the vertex
\begin{equation}\label{veriticesDef}
{\tilde \La}{_{ab}}\!\!\!\!\!{^{cd}}=
^{0\!\!}{\tilde \La}{_{ab}}\!\!\!\!\!{^{cd}}+
^{\la\!\!}{\tilde \La}{_{ab}}\!\!\!\!\!{^{cd}},
\end{equation}
where $^{0\!}{\tilde \La}$ is independent of $\la$, and
$^{\la\!}{\tilde \La}$ vanishes as $e^{-\beta\la}$ in the limit of
$\la\to\infty$. We shall first determine $^{0\!}{\tilde \La}$ and
then, in a second step, $^{\la\!}{\tilde \La}$.

\subsubsection{Voltage induced Particle-Hole Excitations}

\begin{center}
\begin{figure}[t]
\includegraphics[width=6.0cm]{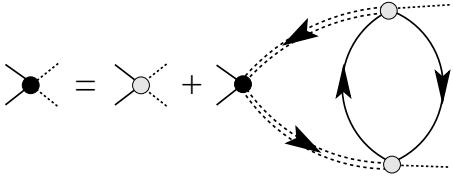}
\caption{\label{fig:vertren}
Vertex equation for the $pf-ce$ vertex, including
only non-logarithmic irreducible parts.}
\end{figure}
\end{center}
We start by discussing the properties of the $ce$ polarization bubble
in Keldysh space, which enters as one building block in the vertex
equation shown in Fig.~\ref{fig:vertren}.  The convolution of two
conduction electron Green functions has the {\it greater} component
\begin{equation}
\ind{\g\!}{\g'\!}{\Pi}{\,\,\,}{\,\,\,>}(\W)=\!\int\!\frac{d\e}{2\pi}
G^{>}_{\g'}(\e+\W)G^{<}_{\g}(\e),
\end{equation}
and in general, the convolution of different Keldysh-components gives
rise to the polarization tensor
\begin{equation}
\ind{\g\!}{\g'\!}{\Pi}{dc}{d'\!c'}(\W)=\!\int\!\frac{d\e}{2\pi}
\undl{G}^{d'\!c'}_{\g'}\!(\e+\W)\undl{G}^{dc}_{\g}(\e).
\end{equation}
It is convenient to form the contraction of this tensor with the
exchange constants $J_{\g\g'}/4$ at each end, and thus define an
effective second order interaction by
\begin{equation}
\ind{}{}{\Pi}{dc}{d'\!c'}(\W)\equiv
\frac{1}{16}J^{2}_{\g\g'}\ind{\g\!}{\g'\!}{\Pi}{dc}{d'\!c'}(\W).
\end{equation}
Contracting again this $ce$ polarization tensor with two bare Keldysh
vertices yields the $pf$ interaction tensor
\begin{eqnarray}
B{_{b'a}}\!\!\!\!\!{^{a'b}}&=&
\La{_{a'b}}\!\!\!\!\!{^{\!c'd}} \Pi{_{dc}}\!\!\!\!\!{^{d'\!c'}}\!
\La{_{ab'}}\!\!\!\!\!{^{\!cd'}}\label{eq:Btensor}\\
&=&\frac{1}{2}\left\{ \Pi^{K}\delta_{a'b}\delta_{ab'}+
    \Pi^{A}\delta_{a'b}\tau^{1}_{ab'}+
    \Pi^{R}\tau^{1}_{a'b}\delta_{ab'} \right\},\nonumber
\end{eqnarray}
where the Langreth rules (cf. Ref.~\onlinecite{Langreth76}) have been
employed to work out the contractions
\begin{equation}
\Pi{_{dc}}\!\!\!\!\!{^{{\ovrl c}{\ovrl d}}}=
2\Pi^{K},\,\,\,\,\,\, \Pi{_{dc}}\!\!\!\!\!{^{c{\ovrl d}}}=
2\Pi^{A},\,\,\,\,\,\, \Pi{_{dc}}\!\!\!\!\!{^{{\ovrl c}d}}=
2\Pi^{R},
\end{equation}
using the notation ${\ovrl 1}=2$ and ${\ovrl 2}=1$ for the Keldysh
indices. As for the single particle Green functions, we organize these
components in a triangular matrix
\begin{equation}
\undl{\Pi}=
\left(\begin{array}{ll}
\Pi^{R} & \Pi^{K} \\
      0 & \Pi^{A}
\end{array}\right),
\end{equation}
and for $\W\ll D$ one finds that
\begin{multline}
\Pi^{R/A}(\W)\!=\!\frac{\pi}{16}g^{2}_{\g\g'}
\left\{\pm(\W+\mu_{\g}-\mu_{\g'})
-i\frac{4D\ln2}{\pi}\right\},\\
\Pi^{K}(\W)\!=\!\frac{\pi}{8}g^{2}_{\g\g'}(\W+\mu_{\g}-\mu_{\g'})
\coth\left(\frac{\W+\mu_{\g}-\mu_{\g'}}{2 T}\right).
\end{multline}
Notice that inter-lead particle-hole excitations do not satisfy the
fluctuation-dissipation theorem as
\begin{equation}
\ind{\g\!}{\g'\!}{\Pi}{\,\,\,}{\,\,\,K}(\W)=
\coth\left(\frac{\W+\mu_{\g}-\mu_{\g'}}{2 T}\right)\!\left[\,
\ind{\g\!}{\g'\!}{\Pi}{\,\,\,}{\,\,\,R}(\W)-
\ind{\g\!}{\g'\!}{\Pi}{\,\,\,}{\,\,\,A}(\W)\right].\nonumber
\end{equation}
The lead-contracted polarization satisfies the following symmetries:
\begin{equation}
\Pi^{>/K/R/A}(-\W)=\Pi^{</K/A/R}(\W),
\end{equation}
and for later use we quote the explicit formula for the ${\it
  greater}$ component, $\Pi^{>}=(\Pi^{K}+\Pi^{R}-\Pi^{A})/2$:
\begin{eqnarray}
\lefteqn{\Pi^{>}(\W)=
\frac{\pi}{8}
\left\{ 
\glr^{2}\left[ (\W+V)(1+N(\W+V))\right.\right.}\\
&&\left.\left.
+(\W-V)(1+N(\W-V))\right]+2\gd^{2}\W(1+N(\W))\right\},\nonumber
\end{eqnarray}
where $N(\W)$ denotes the Bose-function and $\Pi^{>}(\W)=0$ for
$\W\geq 2D+V$. In terms of this function, the second order $pf$ decay
rate may be written as
\begin{equation}
\G_{\al}(\w)=2\theta_{\al\al'}\Pi^{>}(-\w-\al' B/2),\label{eq:pfdamp}
\end{equation}
with $\theta_{\al\al'}=\delta_{\al\al'}+2\tau_{\al\al'}^{1}$.

\subsubsection{Basic Approximations}

The following calculations are based on {\em self-consistent}
perturbation theory to order $g^2$ for self-energies and vertex
corrections. As explained in detail in I, self-consistency is
essential to obtain for example the correct magnetization out of
equilibrium. For this paper, we need a self-consistent resummation of
diagrams to investigate how divergences of bare perturbation theory
are cut off to lowest order in the interactions.
  
However, for non-singular quantities like the lowest-order
self-energy, self-consistency only gives rise to subleading
corrections which we need not keep track of. For example, it is
sufficient to approximate the retarded $pf$ propagators (double-dashed
lines in the diagrams) by
\begin{equation}
{\cal G}^{R}_{\al}(\w)=\frac{1}{\w+\al B/2+i\G_{\al}/2},
\end{equation}
where $\G_{\al}$, given in Eqs.~(\ref{eq:rate1}-\ref{eq:rate4}),
denotes the on-shell decay rate calculated in {\em bare} perturbation
theory. We neglect contributions from $\RE \Sigma_\al(\w)$ which can
be absorbed in a redefinition of $B$ and $g$, and which give rise only
to subleading corrections in the following.
  
To show, formally, that self-consistency does not change this result,
one can use the fact that typical integrals are dominated by
integrations over frequencies in a window of width $\G$ around the
Zeeman levels. Since the various Keldysh components of $\Pi$ vary
slowly with frequency, i.e. $[\Pi(\w+\G)-\Pi(\w)]/\Pi(\w)\sim \G/V$,
we may therefore use $\G/V\sim g^2$ as a small expansion parameter
(this will be shown explicitly for the vertex corrections below).

\subsubsection{Summing up the Ladder}\label{sec:ladder}

To leading order in $\G/V\sim g^2$, the renormalized vertex satisfies
the diagrammatic equation depicted in Fig.~\ref{fig:vertren}. This
equation generates a series of ladder diagrams with dressed $pf$-legs
and bare $ce$ particle-hole propagators as rungs. Contributions from
diagrams with crossed rungs we omit as being of order $\G/V$. In
appendix~\ref{app:crossed}, we establish this relative smallness
explicitly for the crossed 4'th order vertex correction and we expect
higher order corrections to work in the same way.

An iterative solution for the renormalized vertex starts with the
attachment of two $pf$-propagators to the bare Keldysh vertex. This
defines the tensor
\begin{equation}
\ind{\al'}{\al}{V}{ab}{cd}=
2\Lambda^{cd}_{a'b'}{\undl {\cal G}}^{b'\!a}_{\al}
{\undl {\cal G}}^{ba'}_{\al'},
\end{equation}
which is found to have the following components:
\begin{eqnarray}
\ind{\al'}{\al}{V}{11}{cd}&=&
 \delta_{cd}\,\GR_{\al}\GK_{\al'}
+\tau^{1}_{cd}\,\GR_{\al}\GR_{\al'},\\
\ind{\al'}{\al}{V}{12}{cd}&=&
 \delta_{cd}\,\GR_{\al}\GA_{\al'},\\
\ind{\al'}{\al}{V}{21}{cd}&=&
 \delta_{cd}\left\{\GK_{\al}\GK_{\al'}
                 +\GA_{\al}\GR_{\al'}\right\}\nonumber\\
&&\!\!\!\!+\tau^{1}_{cd}\left\{\GK_{\al}\GR_{\al'}
                 +\GA_{\al}\GK_{\al'}\right\},\\
\ind{\al'}{\al}{V}{22}{cd}&=&
 \delta_{cd}\,\GK_{\al}\GA_{\al'}
+\tau^{1}_{cd}\,\GA_{\al}\GA_{\al'}.
\end{eqnarray}
One proceeds by attaching rungs, using the interaction tensor
$B{_{b'\!a}}\!\!\!\!\!{^{a'\!b}}$, and legs consisting of pairs of
dressed $pf$ propagators. This attachment consists of a contraction of
Keldysh, and spin indices, together with an integration over the
frequency circulating the individual sections of the ladder. To
leading order in $\G/V$, we may perform these integrals by neglecting
the slow frequency dependence of the $ce$ polarization functions
compared to the rapid variations in the $pf$ Green functions. Making
use of the identity
\begin{equation}
\frac{1}{a}\frac{1}{b}=
\frac{1}{a-b}\left(\frac{1}{b}-\frac{1}{a}\right),
\end{equation}
products of Green functions may be expressed as either
\begin{eqnarray}
\lefteqn{\GR_{\al}(\W+\w)\GA_{\al'}(\w)=
\frac{1}{\W+(\al-\al')B/2+i(\G_{\al}+\G_{\al'})/2}}\nonumber\\
&&\times
\left(\frac{1}{\w+\al' B/2-i\G_{\al'}/2}-
      \frac{1}{\W+\w+\al B/2+i\G_{\al}/2}\right)\nonumber
\end{eqnarray}
or
\begin{eqnarray}
\lefteqn{\GR_{\al}(\W+\w)\GR_{\al'}(\w)=
\frac{1}{\W+(\al-\al')B/2+i(\G_{\al}-\G_{\al'})/2}}\nonumber\\
&&\times
\left(\frac{1}{\w+\al' B/2+i\G_{\al'}/2}-
      \frac{1}{\W+\w+\al B/2+i\G_{\al}/2}\right)\nonumber
\end{eqnarray}
and likewise for $AR$ and $AA$ products. Considered as an
integral-kernel to be integrated with the various components of the
polarization function, we may neglect the broadening and replace $\W$
by $(\al'-\al)B/2$ inside the parentheses in such products, and
altogether this justifies the approximations
\begin{eqnarray}
\GR_{\al}(\W+\w)\GA_{\al'}(\w)\!&\approx&\!
\frac{2\pi i\delta(\w+\al' B/2)}
     {\W+(\al-\al')B/2+i(\G_{\al}+\G_{\al'})/2},\nonumber\\
\GR_{\al}(\W+\w)\GR_{\al'}(\w)&\approx& 0,\label{eq:approx}
\end{eqnarray}
for a set of legs in the ladder.  Notice that Walker~\cite{Walker68}
has employed a similar approximation in the case of thermal
equilibrium, utilizing the slow frequency dependence of the thermal
$ce$-polarization. In this case, the $RR$ and $AA$ terms are neglected
to leading order in $\G/T$ instead.

Since the legs contain not only retarded and advanced, but also
Keldysh-component Green functions, some of these loop-integrals will
also involve the nonequilibrium $pf$-distribution functions
$n_{\la}(\w)$. This function is found by solving a quantum Boltzmann
equation, obtained as the Keldysh-component of the $pf$ Dyson-equation
with second order $pf$ self-energies. Using the results of I, the
solution at $B=0$ is found to be
\begin{equation}
n_{\la}(\w)=n_{\la}(0)\Pi^{<}(\w)/\Pi^{>}(\w),\label{eq:distrib}
\end{equation}
which, in the case where $\glr\neq 0$ and $T=0$, takes the form
\begin{equation}
n_{\la}(\w)=n_{\la}(0)\left\{\begin{array}{ll}
\frac{\glr^{2}(V-\w)}{(\gll^{2}+\grr^{2})\w+\glr^{2}(V+\w)}
& ,\, 0<\w<V \\ & \\
\frac{\glr^{2}(V-\w)-(\gll^{2}+\grr^{2})\w}{\glr^{2}(V+\w)}
& ,\, -V<\w<0.
\end{array}\right.\label{eq:nfct}
\end{equation}
For $T\to 0$, $n_{\la}(\w)$ vanishes as $e^{-(\w-V)/T}$ for $\w>V$,
and diverges as $e^{-(\w+V)/T}$ for $\w<-V$. For $|\w|<V$,
$n_{\la}(\w)$ crudely resembles a Boltzmann distribution with $T$
replaced by $V/4$. The distribution function clearly inherits the slow
frequency dependence from $\Pi^{>}$ and, to leading order in $\G/V$,
$n_{\la}$ may therefore be treated as a constant, when integrated with
the rapidly varying retarded and advanced $pf$ Green functions. In the
case of $B>0$, the distribution function acquires a spin-index and the
solution is generally more complicated (cf. I).  However, the
frequency dependence is still determined by $\Pi^{>}$, evaluated at
arguments shifted by $\pm B/2$, and therefore remains negligible. In
either case, we are thus allowed to neglect the frequency dependence
of $n_{\la}$, which renders $\GK$ proportional to $\GR-\GA$ by a
constant and reduces all loop-integrals in the ladder to involve only
the products (\ref{eq:approx}) or their complex conjugates.

Omiting all $RR$ and $AA$ terms, $V^{cd}_{ab}$ now simplifies to
\begin{eqnarray}
\ind{\al'}{\al}{V}{11}{cd}&=&
 \delta_{cd}\,M_{\al'\la}\GR_{\al}\GA_{\al'},\\
\ind{\al'}{\al}{V}{12}{cd}&=&
 \delta_{cd}\,\GR_{\al}\GA_{\al'},\\
\ind{\al'}{\al}{V}{21}{cd}&=&
 \delta_{cd}\left\{(1-M_{\al\la}M_{\al'\la})\GA_{\al}\GR_{\al'}
 -M_{\al\la}M_{\al'\la}\GR_{\al}\GA_{\al'}\right\}\nonumber\\
&&\!\!\!\!
+\tau^{1}_{cd}(M_{\al\la}-M_{\al'\la})\GA_{\al}\GR_{\al'},\\
\ind{\al'}{\al}{V}{22}{cd}&=&
 -\delta_{cd}\,M_{\al\la}\GR_{\al}\GA_{\al'},
\end{eqnarray}
and performing the projection $\la\to\infty$, all $pf$-distribution
functions vanish, i.e. $M_{\al\la}\to -1$, and we are left with
\begin{equation}
\ind{\al'}{\al}{V}{ab}{cd}(\W+\w,\w)=
-\delta_{cd}\tau^{3}_{bb}\GR_{\al}(\W+\w)\GA_{\al'}(\w).
\label{eq:simpleV}
\end{equation}

Having performed the projection, it is now a simple matter to sum up
the ladder solving the vertex equation. To keep matters simple we
assume that $B=0$, but once this special case is worked out, a
generalization to $B>0$ will be straightforward. We begin by attaching
the $V$-tensor (\ref{eq:simpleV}) to the $ce$-polarization bubble
defined in (\ref{eq:Btensor}). Working out the contraction, one finds
that
\begin{eqnarray}
\lefteqn{V{_{a'\!b'}}\!\!\!\!\!{^{\!\!\!cd}}\,\,(\W+\w,\w)
B{_{b'\!a}}\!\!\!\!\!{^{a'\!b}}(\w'-\w)=}\\
&&-\delta_{cd}\tau^{3}_{aa}\GR(\W+\w)\GA(\w)\Pi^{>}(\w'-\w).
\end{eqnarray}
We should also attach the Pauli-matrices corresponding to the exchange
vertices at the endpoint vertex and at each end of the polarization
bubble. In zero magnetic field this yields the contraction
\begin{equation}
\tau^{k}_{\al'''\al''}\tau^{i}_{\al''\al}\tau^{j}_{\al'\al'''}
\tau^{i}_{\s\s'}\tau^{j}_{\s'\s}=-2\tau^{k}_{\al'\al},
\label{eq:paulimatrvert}
\end{equation}
which shows that the endpoint $pf$ Pauli-matrix $\tau^{k}$ is carried
through to the new external spin-indices. In this way, the
Pauli-matrix at the endpoint vertex may be left out and the Keldysh
vertex merely receives a factor of $-2$ per rung.

To second order in $g$, the vertex renormalizes to 
\begin{eqnarray}
\lefteqn{^{0\!}{\tilde \La}{_{ab}}\!\!\!\!\!{^{cd}}(\W+\w',\w')}\\
&=&\La{_{ab}}\!\!\!\!\!{^{cd}}-
\int\!\frac{d\w}{2\pi}
V{_{a'\!b'}}\!\!\!\!\!{^{\!\!\!cd}}\,\,(\W+\w,\w)
B{_{b'\!a}}\!\!\!\!\!{^{a'\!b}}(\w'-\w)\nonumber\\
&=&\frac{1}{2}\left\{\tau^{1}_{cd}\delta_{ab}+\delta_{cd}
\left[\tau^{1}_{ab}+i\tau^{3}_{aa}\frac{2\Pi^{>}(\w')}{\W+i\G}\right]
\right\},\nonumber
\end{eqnarray}
where the left superscript $0$ is to remind us that the limit of
$\la\to\infty$ has been taken. The integral over $\w$ is performed
using the $\delta$-function from the $RA$-product of $pf$ Green
functions and $\G$ is the spin-independent ($B=0$) single $pf$
self-energy broadening.

Attaching a set of $pf$ Green functions to this second order
vertex correction, we notice that, after projection and discarding
again all $RR$ and $AA$ products, we have
\begin{equation}
\sum_{a'b'}\delta_{cd}\tau^{3}_{a'a'}{\undl {\cal G}}^{b'a}
{\undl {\cal G}}^{ba'}= -V{_{ab}}\!\!\!\!\!{^{cd}},
\end{equation}
which in turn implies the fourth order correction
\begin{eqnarray}
\lefteqn{^{0\!}{\tilde\La}{_{ab}}\!\!\!\!\!{^{cd}}^{(4)}(\W+\w'',\w'')}
\nonumber\\
&=&2\!\int\!\frac{d\w'}{2\pi}
\frac{i\Pi^{>}(\w')}{\W+i\G}
V{_{\!\!a'b'}}\!\!\!\!{^{\!\!\!c\,d}}\,(\W+\w',\w')
B{_{b'\!a}}\!\!\!\!\!{^{a'\!b}}(\w''-\w')\nonumber\\
&=&\frac{1}{2}\delta_{cd}\tau^{3}_{aa} \frac{2\Pi^{>}(0)}{\W+i\G}
\frac{2\Pi^{>}(\w'')}{\W+i\G}.\label{eq:vertfour}
\end{eqnarray} 
From these two lowest order corrections it is clear how the further
attachment to the ladder will generate a geometric series, and the
vertex function
\begin{equation}
^{0\!}{\tilde \La}{_{ab}}\!\!\!\!\!{^{cd}}(\W+\w,\w)=
\frac{1}{2}\tau^{1}_{cd}\delta_{ab}
+\frac{1}{2}\delta_{cd}\!\left[\tau^{1}_{ab}+\tau^{3}_{aa}
\frac{2i\Pi^{>}(\w)}{\W+i\Gs}\right]
\end{equation}
therefore solves the diagrammatic equation in Fig.\ref{fig:vertren},
in the limit $\la\to\infty$. We employ the suggestive shorthand
\begin{eqnarray}\label{Gamma2}
\Gs&=&\frac{1}{2}(\G_\uparrow+\G_\downarrow)+\G_v\\
   &=&\pi\glr^{2}V \quad \text{for } B,T\ll V,\nonumber
\end{eqnarray}
and, as will be demonstrated in the next section, this is
indeed the spin-relaxation rate. In the present case of zero magnetic
field and isotropic exchange couplings, the longitudinal, and
transverse rates are identical and thus $\Gs=\Gt=\Gl$. In the case of
anisotropic exchange couplings (or in the presence of a finite
magnetic field), spin-flip, and non-spin-flip vertices receive
different corrections and the two rates become discernible. The
anisotropic case will be discussed in Sec.~\ref{anisotropic}.  The
first term in (\ref{Gamma2}) arises from the self-energy,
Eqs.~(\ref{eq:rate1}--\ref{eq:rate4}), the second one,
$\G_v=2\Pi^{>}(0)$, is the vertex correction.  Notice that only
vertices with identical ingoing and outgoing $ce$ Keldysh indices are
renormalized.

So far, we have only determined the $\la\to\infty$ limit of the
vertex, but we need also the second contribution,
$^{\la\!}{\tilde\La}$ in Eq.~(\ref{veriticesDef}), which is
proportional to $e^{-\beta\lambda}$. Having solved for
$^{0\!}{\tilde\La}$ already, we are left with the vertex equation
\begin{eqnarray}
\lefteqn{\hspace*{-2mm}
^{\la\!}{\tilde \La}{_{ab}}\!\!\!\!\!{^{cd}}(\W+\w',\w')=
-2\!\int\!\frac{d\w}{2\pi}
{\tilde \La}{_{a''b''}}\!\!\!\!\!\!\!\!\!\!\!{^{c\,d}}
\,\,\,\,\,(\W+\w,\w)}\\
&&\hspace*{20mm}\times
\undl{\cal G}^{b''\!a'}\!(\W+\w)\undl{\cal G}^{b'\!a''}\!\!(\w)
B{_{b'\!a}}\!\!\!\!\!{^{a'\!b}}(\w'-\w),\nonumber
\end{eqnarray}
which we find to be solved by
\begin{eqnarray}
\lefteqn{^{\la\!}{\tilde \La}{_{ab}}\!\!\!\!\!{^{cd}}(\W+\w,\w)=
-\delta_{cd}\,n_{\la}(0)\left\{
\frac{4\Gs}{\W^{2}+\Gs^{2}}\undl{\Pi}^{ab}(\w)
\right.}\\
&&\left.\!\!
+\frac{i}{\W+i\Gs}\left[
\Pi^{>}(\w)(\tau^{3}_{ab}+i\tau^{2}_{ab})
+\Pi^{<}(\w)(\tau^{3}_{ab}-i\tau^{2}_{ab})
\right]\right\}.\nonumber
\end{eqnarray}
For $c\neq d$ one obtains
\begin{equation}
^{\la\!}{\tilde \La}{_{ab}}\!\!\!\!\!{^{12}}(\W+\w,\w)=
\frac{i[n_{\la}(\W)-n_{\la}(0)]}{\W-i\Gs}\undl{\Pi}^{ab}(\w),
\end{equation}
which is neglected due to the slow frequency dependence of
$n_{\la}(\W)$. It is worth noting, however, that for $B\neq 0$ this
term will in fact be proportional to the magnetization and thus
provide an important renormalization of the $\tau^{1}_{cd}$ term of
the interaction tensor.

This completes the solution of the vertex equation and we may now
proceed to determine its influence on physical observables. In doing
so, one has to attach a pair of $pf$ Green functions to the
renormalized vertex, and most often one may therefore continue to use
the approximation (\ref{eq:approx}).  Since the dependence of the
vertex on the relative frequency $\w$ is set by $\undl{\Pi}^{ab}(\w)$,
one can safely set $\w$ to $0$ and consider the vertex as a function
of $\W$ alone. With $\G_{v}=2\Pi^{>}(0)$, the renormalized vertex then
simplifies to
\begin{equation}
{\tilde\La}{_{ab}}\!\!\!\!\!{^{cd}}(\W)=
\frac{1}{2}\tau^{1}_{cd}\delta_{ab}
+\frac{1}{2}\delta_{cd}L_{ab}(\W),\label{eq:cmvertex}
\end{equation}
where $L_{ab}=^{0\!}L_{ab}+^{\la\!}L_{ab}$, with
\begin{equation}
^{0\!}L_{ab}(\W)=\tau^{1}_{ab}+\tau^{3}_{aa}
\frac{i\G_{v}}{\W+i\Gs}=
\left(\begin{array}{cc}
  \frac{i\G_{v}}{\W+i\Gs} & \frac{\W+i(\G+2\G_{v})}{\W+i\Gs}\\ & \\
  \frac{\W+i\G}{\W+i\Gs} & -\frac{i\G_{v}}{\W+i\Gs}
\end{array}\right)_{\!\!ab}
\end{equation}
and 
\begin{equation}
^{\la\!}L_{ab}(\W)=-2n_{\la}(0)\left[
\frac{4\Gs\,\undl{\Pi}^{ab}(0)}{\W^{2}+\Gs^{2}}+\tau^{3}_{ab}
\frac{i\G_{v}}{\W+i\Gs}\right],
\end{equation}
where $n_{\la}(0)\propto e^{-\beta \lambda}$. Using this result we can
now calculate physical quantities like susceptibility and T-matrix.

\begin{center}
\begin{figure}[t]
\includegraphics[width=3.5cm]{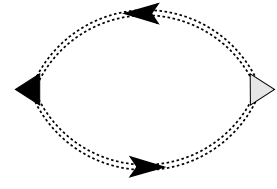}
\caption{\label{fig:suscept}
  Dynamical susceptibility. Triangles refer to external measurement
  vertices. The black (emission) vertex is renormalized like the
  interaction vertex in Fig.~\ref{fig:vertren}, except that the two
  external $ce$ legs are removed. The other (absorption) vertex
  remains undressed.}
\end{figure}
\end{center}

\subsection{Dynamical spin susceptibility}\label{susceptibility}

In order to uncover the physical meaning of the rate $\Gs$ introduced
in Eq.~(\ref{Gamma2}), we include here a brief discussion of the
transverse spin susceptibility:
\begin{equation}
{^{\perp}}\chi^{R}(t)=i\theta(t)\langle[S^{-}(t),S^{+}(0)]\rangle.
\end{equation}
The {\it transverse} spin relaxation rate, $\Gt$, is defined as the
broadening of the resonance pole in this response function, and as
will be shown below, $\Gs$ plays exactly this role. Throughout this
Section, we may therefore use $\Gs=\Gt$. With a suitable
generalization of $\Gv$, entering Eq. (\ref{Gamma2}), which will be
given in Sec.~\ref{anisotropic}, this identification holds also for
anisotropic coupling.

Translating to the pseudo fermion representation on the Keldysh
contour, the transverse susceptibility is calculated from
\begin{equation}
{^{\perp}}\chi(\tau)=-i(-i)^{2}\langle T_{c_{K}}\{
f^{\dagger}_{\down}(\tau)f_{\up}(\tau)
f^{\dagger}_{\up}(0)f_{\down}(0)\}\rangle,
\end{equation}
which in turn leads to the Feynman diagram in Fig.~\ref{fig:suscept}
when including vertex, and $pf$ self-energy corrections. The bare
absorption, and emission vertices are given as
$\al^{1}_{ab}=\frac{1}{\sqrt{2}}\delta_{ab}$ and
${\tilde\al}^{1}_{ab}=\frac{1}{\sqrt{2}}\tau^{1}_{ab}$, respectively
(cf.~I). The absorption vertex is kept undressed and the emission
vertex renormalizes like the interaction vertex-component
$\sqrt{2}{\tilde \La}{_{ab}}\!\!\!\!\!{^{11}}$, whereby
\begin{equation}
{^{\perp}}\chi^{R}(\W)=i\int\frac{d\w}{2\pi}
{\tilde\La}{_{ab}}\!\!\!\!\!{^{11}}(\W+\w,\w)
\undl{\cal G}_{\up}^{bc}(\W+\w)\undl{\cal G}_{\down}^{ca}(\w).
\end{equation}
Notice that the canonical ensemble average, enforcing single occupancy
on the dot, is carried out by dividing the $\la$-dependent
grand-canonical average by $\langle Q\rangle_{\la}$ and taking the
limit $\la\to\infty$ (cf.~I). This procedure affects only the $pf$
distribution functions and allows to neglect all terms proportional to
squares, or higher powers of $n_{\al\la}$. Working out the
contractions, we arrive at
\begin{widetext}
\begin{eqnarray}\label{eq:rensusc}
{^{\perp}}\chi^{R}(\W)&=&i\int\frac{d\w}{2\pi}\Bigl\{\,\,
\!\!\ ^{0\!}{\tilde \La}{_{21}}\!\!\!\!\!{^{11}}(\W+\w,\w)
2\left[n_{\down}(\w)-n_{\up}(\W+\w)\right]
\GR_{\up}(\W+\w)\GA_{\down}(\w)
\nonumber\\
&&\hspace{12mm}+\left[
 {^{\la\!}}{\tilde \La}{_{11}}\!\!\!\!\!{^{11}}(\W+\w,\w)
+{^{\la\!}}{\tilde \La}{_{21}}\!\!\!\!\!{^{11}}(\W+\w,\w)
-2n_{\down}(\w){^{0\!}}{\tilde \La}{_{21}}\!\!\!\!\!{^{11}}(\W+\w,\w)
\right]\GR_{\up}(\W+\w)\GR_{\down}(\w)
\nonumber\\
&&\hspace{12mm}+\left[
 {^{\la\!}}{\tilde \La}{_{22}}\!\!\!\!\!{^{11}}(\W+\w,\w)
-{^{\la\!}}{\tilde \La}{_{21}}\!\!\!\!\!{^{11}}(\W+\w,\w)
+2n_{\up}(\W+\w){^{0\!}}{\tilde \La}{_{21}}\!\!\!\!\!{^{11}}(\W+\w,\w)
\right]\GA_{\up}(\W+\w)\GA_{\down}(\w)
\Bigr\}.
\end{eqnarray}
\end{widetext}
The important fact that the final result is proportional to $n_{\al}$
is ensured by the relations
\begin{equation}
{^{0\!}}{\tilde \La}{_{21}}\!\!\!\!\!{^{11}}
+{^{0\!}}{\tilde \La}{_{11}}\!\!\!\!\!{^{11}}=
{^{0\!}}{\tilde \La}{_{21}}\!\!\!\!\!{^{11}}
-{^{0\!}}{\tilde \La}{_{22}}\!\!\!\!\!{^{11}}=1,
\end{equation}
as such a constant drops after integrating over $\GR\GR$ or $\GA\GA$.

In the limit of $B\to 0$, the factor of $n_{\down}(\w)-n_{\up}(\W+\w)$
in the first term is of order $\W/V$, and therefore we are forced to
keep also the other terms involving $\GR\GR$ or $\GA\GA$. In this
case, we have to keep the full dependence of the vertex on two
frequencies, but since for example the parts of the vertex which are
retarded with respect to $\w$ integrate to zero with $\GR\GR$, matters
simplify substantially. The first square bracket can simply be
replaced by
\begin{equation}
\left[\frac{2n_{\la}(0)\Pi^{<}(\w)-2\Pi^{>}(\w)n_{\la}(\w)}{\W+i\Gt}
-in_{\la}(\w)\right],\label{eq:cancel}
\end{equation}
and inserting now the {\it nonequilibrium} distribution function given
by (\ref{eq:distrib}), the first two terms of this expression are seen
to cancel.  We emphasize the fact that this important cancellation
takes place only when using the correct distribution function, i.e.
the solution to the quantum Boltzmann equation corresponding to second
order $pf$ self-energies.

The term involving $\GA\GA$ works in a similar way, and using the
approximation ${\cal G}^{R/A}_{\up}(\W+\w){\cal
  G}^{R/A}_{\down}(\w)\approx -\partial_{\w}(\w\pm i0_{+})^{-1}$,
valid to leading order in $\max(|\W|,\G)/V$ when integrated with the
slowly varying distribution function, the last two terms in
(\ref{eq:rensusc}) may be evaluated by partial integration. The first
term comes with a factor of ${^{0\!}}{\tilde
  \La}{_{21}}\!\!\!\!\!{^{11}}(\W+\w,\w)
\GR_{\up}(\W+\w)\GA_{\down}(\w)\approx 2\pi i\delta(\w)/(\W+i\Gt)$,
and altogether one finds that
\begin{equation}
{^{\perp}}\chi^{R}(\W)
\approx\frac{M}{B}\frac{i\Gt}{\W+i\Gt},\label{eq:chiR0}
\end{equation}
for $\max(|\W|,\G)\ll V$. The prefactor is independent of $B$ and is
obtained as the derivative $-n'(0)$, with $n(\w)$ given by
(\ref{eq:nfct}) and with the replacement $n_{\la}(0)\to 1/2$, due to
the normalization by $\langle Q\rangle_{\la}$ before projection. The
zero-frequency limit obeys ${^{\perp}}\chi^{R}(0)=M/B$, like
in equilibrium, and the nonequilibrium magnetization was found in I to
be
\begin{equation}
M=\frac{(\gll^{2}+\grr^{2}+2\glr^{2})B}{2\glr^{2}V},
\end{equation}
similar to a Curie-law with $1/T$ replaced by $4/V$. Notice that the
result ($\ref{eq:chiR0}$), has been obtained also in
Ref.~\onlinecite{Mao03}, using a Majorana-fermion representation.

In the case of a finite magnetic field, the factor of
$n_{\down}(\w)-n_{\up}(\W+\w)$ in the first term of (\ref{eq:rensusc})
will be of order $B/V$. For $B\gg\max(|\W+B|,\G)$, this term will
therefore dominate the other terms involving $\GR\GR$ or $\GA\GA$. For
$B>0$, the vertex renormalization is modified, but since we only need
to consider the first term in (\ref{eq:rensusc}), only a single
component is needed. For this particular component the generalization
is straightforward and one finds that
\begin{equation}
{^{0\!}}{\tilde \La}{_{21}}\!\!\!\!\!{^{11}}(\W+\w,\w)=
\frac{1}{2}\left(
1-\frac{2i\Pi^{>}(\w-B/2)}{\W+B+i\Gt}
\right),
\end{equation}
where $\Gt$ is given in Eq.~(\ref{Gamma2}) and 
depends now on both $V$ and $B$ (see Eq.~(\ref{gammaFormula}) below). 
The integral over $\w$ is performed using the
approximation (\ref{eq:approx}), and the susceptibility is found to be
\begin{equation}
{^{\perp}}\chi^{R}(\W)\approx\frac{M}{\W+B+i\Gt},\label{eq:chiR}
\end{equation}
valid for $\max(|\W+B|,\G)\ll\min(B,V)$

In the intermediate regime where $B\ll\min(|\W+B|,\G)$, one would need
to generalize also the $\la$-dependent part of the vertex to the case
of $B>0$.  However, we expect that cancellations, similar to those
found in terms like (\ref{eq:cancel}) at zero field, will take place
also at finite $B$, once the correct $B$-dependent distribution
function is used. In this manner, we expect the general formula for the
susceptibility to be simply
\begin{equation}
{^{\perp}}\chi^{R}(\W)\approx\frac{M}{B}\frac{B+i\Gt}{\W+B+i\Gt},
\end{equation}
valid for $\max(|\W+B|,\G)\ll V$. This function obviously has the
correct asymptotic behaviors, corresponding to (\ref{eq:chiR0}) and
(\ref{eq:chiR}), and is consistent with the equilibrium
result\cite{Walker68,Woelfle71}.

For completeness, we state here the relevant asymptotics of $\Gt$ as a
function of $V$, $B$ and $T$:
\begin{equation}
\Gt\approx\left\{\begin{array}{ll}
\pi\glr^{2}V & ,\,\max(T,B)\ll V\\ & \\
\pi(\gll^{2}+\grr^{2})B/4 & ,\,\max(T,V)\ll B\\ & \\
\pi(\gll^{2}+\grr^{2}+2\glr^{2})T & ,\,\max(B,V)\ll T
\end{array}\right. \label{gammaFormula}
\end{equation}
In the equilibrium limit, $V=0$, this corresponds to the result
obtained in Refs.~\onlinecite{Walker68} and \onlinecite{Woelfle71},
$\Gt\approx\pi g^{2}\max(T,B/4)$.

\section{Conduction electron T-matrix}\label{tmatrix}

With the renormalized vertex at hand, we now proceed to calculate the
conduction electron T-matrix, including the leading logarithmic
corrections. The T-matrix, $T_{\alpha \alpha'}$, is of great physical
significance, insofar as it describes the scattering of conduction
electrons from lead $\alpha'$ to lead $\alpha$, and thereby also the
transport across the dot. It is determined from the conduction
electron Green function:
\begin{equation}
G^{R}_{\g\g',\s}(\w)=G^{R(0)}_{\g\s}(\w)\delta_{\g\g'}+
G^{R(0)}_{\g\s}(\w){\rm T}^{R}_{\g\g',\s}(\w)G^{R(0)}_{\g'\s}(\w).
\end{equation}
In cases where the exchange-tunneling Hamiltonian (\ref{eq:hamilton})
is derived from an underlying Anderson model, i.e. from a single
quantum dot in the Coulomb-blockade regime (cf. e.g.
Ref.~\onlinecite{Kaminski99}), one has $J_{LR}^2=J_{LL} J_{RR}$ and
only one of the eigenvalues of the $2\times 2$ matrix $T_{\g\g'}$ is
finite. In such a
\begin{center}
\begin{figure}[ht]
\includegraphics[width=0.95\linewidth]{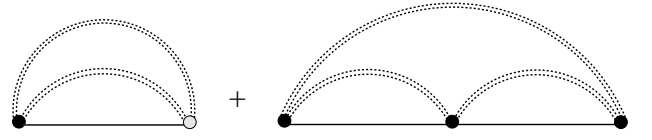}
\caption{\label{fig:tmatr}
  Diagrams for the conduction electron T-matrix, with dressed $pf$
  propagators and dressed interaction vertices (black dots).}
\end{figure}
\end{center}
situation, $\IM[{\rm T}_{\g\g'\s}(\w)]$ is, at low energies, directly
proportional to the spectral function of the electrons on the dot (see
e.g.  Ref.~\onlinecite{Rosch03b} and references therein). This
spectral function can be measured directly by tunneling into the
dot\cite{Franceschi02}, and henceforth we shall focus on the imaginary
part of $T_{\alpha \alpha'}$.

In Fig.~\ref{fig:tmatr} we show the two diagrams contributing to the
T-matrix to third order. Within bare perturbation theory (i.e. using
bare vertices and Green functions in Fig.~\ref{fig:tmatr}), one
obtains the following intra- and inter-lead components at $T,B=0$:
\begin{widetext}
\begin{eqnarray}\label{pert1}
\IM\left[{\rm T}_{\g\g}^{R}(\W)\right]&=&-\frac{3\pi}{16 \Nf}
\left\{(g_{\g\g}^{2}+\glr^{2})
\left[1+2g_{\g\g}\ln\left(\frac{D}{|\W-\mu_{\g}|}\right)\right]
+4\gd\glr^{2}\ln\left(\frac{D}{|\W+\mu_{\g}|}\right)
\right\},\\
\IM\left[{\rm T}_{LR}^{R}(\W)\right]&=&-\frac{3\pi}{16 \Nf}\glr
\left\{
2\gd\left[1+2g_{LL}\ln\left(\frac{D}{|\W-\mu_{\g}|}\right)\right]
+2 (g_{RR}^{2}+\glr^{2})\ln\left(\frac{D}{|\W-\mu_{R}|}\right)
\right\},\label{eq:pertselfen}
\end{eqnarray}
\end{widetext}
with $\mu_L=-\mu_R=V/2$. Within bare perturbation theory, the T-matrix
diverges close to each Fermi surface, or more precisely, for $\W \to
\mu_\g$, some of the logarithms are cut off by the voltage
$V=\mu_L-\mu_R$ while others remains unaffected. In this sense
voltage and temperature act very differently as $T$ would cut off all
logarithmic terms uniformly.  The central question formulated in the
introduction is, how the logarithmic divergences which remain for $T
\to 0$ and large $V$ are cut off when the perturbation theory is
properly resummed. To find the correct cut-off to order $g^2$, we have
to replace the bare Green functions and bare vertices in
Fig.~\ref{fig:tmatr} by the dressed ones.

As the second-order diagram in Fig.~\ref{fig:tmatr} gives only a
finite contribution $\IM[{\rm
  T}_{\g\g'}^{R}(\W)]=-\frac{3\pi}{16\Nf}\sum_{\g''} g_{\g \g''}
g_{\g''\g'}$, the inclusion of self-energy, and vertex corrections
will produce only subleading corrections of order $g^4$, as can be
shown by an explicit calculation.

The fate of the logarithms arising to order $g^3$ is more interesting,
and in the following we will therefore carefully evaluate the second
diagram in Fig.~\ref{fig:tmatr}. This contribution involves the
spin-contractions
\begin{equation}
\tau^{k}_{\al\al'}\tau^{j}_{\al'\al''}\tau^{i}_{\al''\al}
\tau^{i}_{\s\s''}\tau^{j}_{\s''\s'}\tau^{k}_{\s'\s}=24  
\label{eq:paulmatrTp}
\end{equation}
 for the Peierls, and
\begin{equation}
\tau^{i}_{\al\al''}\tau^{j}_{\al''\al'}\tau^{k}_{\al'\al}
\tau^{i}_{\s\s''}\tau^{j}_{\s''\s'}\tau^{k}_{\s'\s}=-24
\label{eq:paulmatrTc}
\end{equation}
for the Cooper-channel. Writing out the sum of these two types of
diagrams, corresponding to different orientations of the $pf$-loop,
one finds that
\begin{widetext}
\begin{eqnarray}
{\rm T}_{\g\g'}^{R(3)}(\W)&=&
\frac{24}{2}(-1)^{1}(i)^{3}(J_{\g\g''}J_{\g''\g'''}J_{\g'''\g'}/4^{3})
\!\int\!\frac{d\w}{2\pi}\!\int\!\frac{d\e}{2\pi}\!\int\!
\frac{d\e'}{2\pi}{\undl G}_{\g''}^{d''\!c''}(\W+\e)
{\undl G}_{\g'''}^{d'\!c'}(\W+\e')\nonumber\\
&&\hspace*{3mm}\times\left\{
{\tilde\La}^{1d''}_{a''\!b}(\w,\w+\e){\undl {\cal G}}^{ba}(\w)
{\tilde\La}^{c'\!1}_{ab'}(\w+\e',\w){\undl {\cal G}}^{b'\!a'}(\w+\e')
{\tilde \La}^{c''\!d'}_{a'\!b''}(\w+\e,\w+\e')
{\undl {\cal G}}^{b''\!a''}(\w+\e)
\right.\nonumber\\
&&\left.\hspace*{5.5mm}
-{\tilde\La}^{c'\!1}_{a'\!b}(\w,\w-\e'){\undl {\cal G}}^{ba}(\w)
{\tilde\La}^{1d''}_{ab''}(\w-\e,\w){\undl {\cal G}}^{b''\!a''}(\w-\e)
{\tilde\La}^{c''\!d'}_{a''\!b'}(\w-\e',\w-\e)
{\undl {\cal G}}^{b'\!a'}(\w-\e') \right\}.\label{eq:cese3}
\end{eqnarray}
\end{widetext}
This Keldysh contraction has a total of 256 terms, of which only a few
will contribute in the end. Some will involve a ${\undl {\cal
    G}}^{21}$, which is zero, and others will involve a product of
more than one lesser-component $\GL$, which, being proportional to
higher powers of the $pf$ distribution function, will vanish faster
than $\langle Q\rangle_{\la}$. Since the Keldysh representation
contains $\GL$ as part of $\GK=\GL+\GG$, it is a daunting task to
isolate all contractions with only one factor of $\GL$. Nevertheless,
since we are dealing here with a trace over the $pf$ Keldysh indices,
we are free to work in a more convenient basis for the pseudo
fermions. Thus choosing
\begin{equation}
\zeta=\left(\begin{array}{rr}
1 & 1 \\
0 & 1
\end{array}\right),
\hspace*{5mm}
\zeta^{-1}=\left(\begin{array}{rr}
1 & -1 \\
0 &  1
\end{array}\right),
\end{equation}
the Keldysh matrix Green functions may be transformed as
\begin{equation}
\Gpf=\zeta\,{\undl {\cal G}}\,\zeta^{-1}=
\left(\begin{array}{cc}
\GR & 2\GL \\
  0 & \GA
\end{array}\right),
\end{equation}
which has the nice property that $\Gpf$ becomes diagonal after
projection. The renormalized vertices may be considered as functions
of only one frequency and therefore take the form (\ref{eq:cmvertex}),
which we write loosely as $\La=\delta+L$. For opposite $ce$
Keldysh-indices the vertex retains the structure of the
identity-matrix $\delta$ under the transformation. For equal
$ce$-indices, the matrix $L_{ab}$ transforms to
\begin{equation}
\Lpf=\zeta L\,\zeta^{-1}
    ={^{0\!}}\Lpf+{^{\la\!}}\Lpf,
\end{equation}
where
\begin{equation}
^{0\!}\Lpf=
\left(\begin{array}{cc}
   1 & 0 \\
\phi & -1
\end{array}\right),
\end{equation}
and
\begin{equation}
^{\la\!}\Lpf=2n_{\la}(0)
\left(\begin{array}{cc}
\frac{\psi^{\ast}}{2}-\frac{4\Gs\Pi^{R}(0)}{\W^{2}+\Gs^{2}} & 
\psi \\
0 &
-\frac{\psi^{\ast}}{2}-\frac{4\Gs\Pi^{A}(0)}{\W^{2}+\Gs^{2}}
\end{array}\right),
\end{equation}
with $\phi=\frac{\W+i\G}{\W+i\Gs}$ and $\psi=\frac{2i\G_{v}}{\W-i\Gs}$.

In this representation the contraction in Eq. (\ref{eq:cese3}) becomes
manageable and one has to deal with merely 8 different types of terms.
The full contraction is worked out in Appendix~\ref{app:contract},
resulting in
\begin{eqnarray}
\lefteqn{\hspace*{-1mm}{\rm T}_{\g\g'}^{R(3)}(\W)=
\frac{3}{16}n_{\la}J^{3}_{\g\g''\g'''\g'}
\!\int\!\frac{d\e}{2\pi}\!\int\!\frac{d\e'}{2\pi}}\label{eq:imcese3}\\
&&\times\left\{G_{\g''}^{K}(\W+\e)G_{\g'''}^{R}(\W+\e')
\GA_{\Gs}(\e)\GA_{\Gs}(\e')\right.\nonumber\\
&&\hspace*{4mm}
-\left[G_{\g''}^{R}(\W+\e)G_{\g'''}^{K}(\W+\e')\right.\nonumber\\
&&\left.\left.\hspace*{7.5mm}
+G_{\g''}^{K}(\W+\e)G_{\g'''}^{A}(\W+\e')\right]
\GA_{\Gs}(\e)\GA_{\Gs}(\e-\e')\right\},\nonumber
\end{eqnarray}
where $\GA_{\Gs}(\e)=1/(\e-i \Gs)$ are Green functions broadened by
$\Gs$ rather than $\G/2$ and we use the shorthand notation
$J^{3}_{\g\g''\g'''\g'}=J_{\g\g''}J_{\g''\g'''}J_{\g'''\g'}$. Already
at this stage, it is apparent that the vertex corrections have served
to replace twice the $pf$ self-energy broadening by $\Gs$. Making use
of the basic integrals,
\begin{equation}
\int_{-D}^{D}\!d\e\, \frac{{\rm
    sgn}(\e+a)(\e+b)}{(\e+b)^{2}+\Gs^{2}}=
\ln\left(\frac{D^{2}}{(b-a)^{2}+\Gs^{2}}\right)
\end{equation}
and
\begin{equation}
\int_{-D}^{D}\!d\e\, \frac{{\rm sgn}(\e+a)\G}{(\e+b)^{2}+\Gs^{2}}=
2\tan^{-1}\left(\frac{b-a}{\Gs}\right),
\end{equation}
representing a broadened logarithm and a broadened sign-function,
respectively, the remaining integrals over $\e$ and $\e'$ are
straightforward.

The first line of the integral (\ref{eq:imcese3}) involves a
convolution of $G^{K}$ with $\GA$, which yields
\begin{equation}
-i\Nf\left\{\ln\left(\frac{D^{2}}{(\W-\mu_{\g''})^{2}+\Gs^{2}}\right)
  +2 i\tan^{-1}\left(\frac{\W-\mu_{\g''}}{D}\right)\right\}.\nonumber
\end{equation}
This term is multiplied by the convolution of $G_{\g'}^{R}$ with
$\GA_{\G}$, equal to $i G_{\g'}^{R}(\W+i\G)$, and altogether the first
line yields the imaginary part
\begin{equation}
-2 n_{\la}\frac{3\pi}{32}J^{3}_{\g\g''\g'''\g'}\Nf^{2}
\ln\left(\frac{D^{2}}{(\W-\mu_{\g''})^{2}+\Gs^{2}}\right).
\label{eq:line1}
\end{equation}

Using a spectral representation for the $ce$ Green functions, the
remaining two lines of (\ref{eq:imcese3}) can be brought to the form
\begin{eqnarray}
\lefteqn{\hspace*{-4mm}
2n_{\la}\frac{3}{32}J^{3}_{\g\g''\g'''\g'}\Nf^{2}
\!\int_{-D}^{D}\!\frac{d\e}{2\pi}\,
\!\int_{-D}^{D}\!d\w\,}\nonumber
\\
&&\hspace*{16mm}\times\!
\left[\,\frac{{\rm sgn}(\e-\mu_{\g'''})}{(\w-\W-i\Gs)(\w-\e-i\Gs)}
\right.\nonumber\\
&&\hspace*{18mm}\left.
+\frac{{\rm sgn}(\e-\mu_{\g''})}{(\e-\W-i\Gs)(\w-\e+i\Gs)}\,\right].
\end{eqnarray}
The $\w$ integral in first term vanishes in the limit $D\to\infty$,
and keeping $D$ finite this term remains smaller than the second term
by a factor of $\W/D$ or $\Gs/D$. Keeping only the second term, the
imaginary part takes exactly the same form as (\ref{eq:line1}), and
finally we obtain after projection
\begin{widetext}
\begin{eqnarray}
\IM\left[{\rm T}_{\g\g}^{R}(\W)\right]=-\frac{3\pi}{16 \Nf}
\left\{(g_{\g\g}^{2}+\glr^{2})\left[1+g_{\g\g}
\ln\left(\frac{D^{2}}{(\W-\mu_{\g})^{2}+\Gs^{2}}\right)\right]
+2\glr^{2}\gd\ln\left(\frac{D^{2}}{(\W+\mu_{\g})^{2}+\Gs^{2}}\right)
\right\},\label{eq:finalselfen}
\end{eqnarray}
\end{widetext}
with no summation over $\g$ implied. We find precisely the result of
bare perturbation theory, Eq.~(\ref{pert1}), but now with the
logarithmic divergences cut off by $\Gs$. The same conclusion also
holds for ${\rm T}_{LR}$. This is the central result of this paper.

\section{Anisotropic Couplings: $T_1$ vs. $T_2$}\label{anisotropic}

The longitudinal, and the transverse spin-relaxation rates, $1/T_1$
and $1/T_2$, have rather different physical interpretations. It is
therefore interesting to determine which combination of the two rates
actually controls the logarithmic divergences. In the previous chapter
we restricted ourselves to the case of zero magnetic field and
isotropic couplings, and in this case we cannot distinguish between the
two rates as $1/T_1=1/T_2=\Gs$.

To discriminate between the two rates, even for $B=0$, we generalize
the exchange interaction to involve two different couplings
\begin{equation}
 {^{\bot\!\!}}J_{\g'\g}(\tau^{1}_{\al'\al}\tau^{1}_{\s'\s}
                      +\tau^{2}_{\al'\al}\tau^{2}_{\s'\s})
+{^{z\!\!}}J_{\g'\g}\tau^{3}_{\al'\al}\tau^{3}_{\s'\s},
\label{eq:anisexch}
\end{equation}
and we may now repeat all calculations above, keeping track of
separate spin-flip and non-spin-flip processes. Since we consider only
the case of zero magnetic field, the $pf$ self-energy broadening
remains spin-independent and we obtain from Eq. (\ref{eq:pfdamp}), for
$V\gg T$ and $B=0$,
\begin{equation}
\G=\G_\uparrow=\G_\downarrow=
\frac{\pi}{4}({^{z\!}}\glr^{2}+2\,{^{\bot\!}}\glr^{2})V.
\end{equation}

The vertex corrections now take a different form, depending on whether
or not the spin is flipped at the vertex. For $T=B=0$ and finite $V$
we obtain $\Gv^\perp=\pi\,{^{z\!}}\glr^{2}V/4$ for the spin-flip
vertex and $\Gv^z=\pi(2\,{^{\bot\!}}\glr^{2}-\,{^{z\!}}\glr^{2})V/4$
in the case of no spin-flip.  Therefore, the longitudinal, and the
transverse spin-relaxation rates are given  by
\begin{equation}
\frac{1}{T_1}=\Gl=\G+\Gv^z=\pi\,{^{\bot\!}}\glr^{2}V,
\end{equation}
and
\begin{equation}\label{gamma2Aniso}
\frac{1}{T_2}=\Gt=\G+\Gv^\perp=
\frac{\pi}{2}({^{z\!}}\glr^{2}+{^{\bot\!}}\glr^{2})V.
\end{equation}
Notice that $1/T_1=0$ for $^\perp\!g=0$. This is due to a cancellation
of vertex, and self-energy corrections, reflecting the conservation of
$S_{z}$ in this case.
 
How do these spin-relaxation rates modify the logarithmic divergences?
A close inspection of the Keldysh contractions and the integrals
carried out in Appendix~\ref{app:contract} reveals that only the
${^{0\!}}\La$-part of the renormalized vertex connecting to the
out-going $ce$-line (i.e. the left most vertex in
Fig.~\ref{fig:tmatr}) gives rise to a logarithmic divergence, and
furthermore determines whether this logarithm is cut off by $\Gt$
or $\Gl$ depending on whether this vertex involves a spin-flip or not.
Therefore, in the case of anisotropic couplings,
Eq.~(\ref{eq:finalselfen}) generalizes to
\begin{widetext}
\begin{eqnarray}\label{anisotropicT}
\IM\left[{\rm T}_{\g\g'}^{R}(\W)\right]&=&-\frac{\pi}{16\Nf}
\sum_{\g'',\g'''}
\left\{\,\,\,
{^{z\!}}g_{\g\g''}
\left[{^{z\!}}g_{\g''\g'}
+{^{\bot\!}}g_{\g''\g'''}{^{\bot\!}}g_{\g'''\g'}
\ln\left(\frac{D^{2}}{(\W-\mu_{\g''})^{2}+\Gl^{2}}\right)
\right]\right.\\
&&\hspace{25.15mm}\left.
+
{^{\bot\!}}g_{\g\g''}
\left[2\,{^{\bot\!}}g_{\g''\g'}+
\left({^{z\!}}g_{\g''\g'''}{^{\bot\!}}g_{\g'''\g'}+
      {^{\bot\!}}g_{\g''\g'''}{^{z\!}}g_{\g'''\g'}\right)
\ln\left(\frac{D^{2}}{(\W-\mu_{\g''})^{2}+\Gt^{2}}\right)
\right]
\right\}.\nonumber
\end{eqnarray}
\end{widetext}
Roughly speaking, two thirds of the logarithms are broadened by $\Gt$
and one third by $\Gl$.  

How are these results modified beyond lowest order perturbation
theory? In Appendix~\ref{app:xray} we investigate this question in the
limit $^\perp\!g\to 0$ for {\em finite} $^z\!g$. In this limit, the
logarithmic singularities in correlation functions like $\langle
S^{-}S^{+}\rangle$ resum in equilibrium to power-laws with
exponents depending on $^z\!g$. In Appendix~\ref{app:xray} we use a
mapping of our problem to a non-equilibrium X-ray edge problem
together with results by Ng\cite{Ng96} and
others\cite{Combescot00,Muzykantskii03} to investigate how these
power-law singularities are affected by a finite bias voltage and the
associated current. We find that all these power-laws are cut off by a
rate related to $1/T_2$.  This has a simple interpretation: for finite
$^z\!\!J$ a finite current is flowing through the system and the
corresponding noise prohibits the coherence of the two external
spin-flips at low energy. Close inspection reveals, that the second
logarithm in Eq.~(\ref{anisotropicT}) is calculated from a correlation
function of the type discussed in Appendix~\ref{app:xray}.  The
non-perturbative results of the Appendix therefore confirm our
perturbative Eq.~(\ref{anisotropicT}). The first logarithm in
Eq.~(\ref{anisotropicT}), however, arises from a different correlator
(as one external vertex involves $S_z$) which we have not tried to
calculate to higher orders in $^z\!g$.

Also in the presence of a magnetic field the situation is more
complex and at present we do not know which combination of relaxation
rates controls the logarithmic divergences arising for $V\approx B$.
The vertex corrections depend on $B$ and, as mentioned in
Sec.~\ref{sec:ladder}, also the $\tau^{1}_{cd}$-part of the vertex
renormalizes in this case. Furthermore, the non-spin-flip vertex
depends on the orientation of the incoming spin, and its two different
components are found only after solving two coupled vertex equations
(cf. e.g. Ref.~\onlinecite{Walker68}).

In many physical situations, $1/T_1$ and $1/T_2$ differ only by a
numerical prefactor of order $1$ and such a factor in the argument of
the logarithms is not important. In this situations it is not
necessary to keep track of differences of $1/T_1$ and $1/T_2$, if one
is interested in a calculation to leading order in
$1/\ln[\max(V,B)/T_K]$ (cf. e.g. Ref.~\onlinecite{Rosch03a}).

\section{Discussion}

In this paper we have addressed the question how, far out of
equilibrium, the presence of a sufficiently large current prohibits
the coherent spin-flips necessary for the development of the Kondo
effect. In an explicit calculation, we have confirmed the expected
answer~\cite{Wolf69,Meir93,Kaminski99,Rosch01,Rosch03a} that the
spin-relaxation rate cuts off the logarithmic corrections of
perturbation theory. This implies that for $\Gs \gg T_K$ (i.e. for
$V\gg T_K$, see Refs.~\onlinecite{Rosch01,Rosch03a}), the Kondo-model
stays in the perturbative regime, which allows calculating its
properties in a controlled way using perturbative renormalization
group~\cite{Rosch03a}.

We have worked out this scenario explicitly for the imaginary part of
the conduction electron T-matrix, taking into account the joint effect
of self-energy, and vertex corrections.  In the limit of zero
temperature and $\ln(V/\TK)\gg 1$, perturbation theory remains valid
and the vertex corrections were determined by summing up diagrams to
leading order in $\G/V\sim g^2$. Within bare perturbation theory, the
T-matrix exhibits logarithmic divergences at the Fermi energies of the
left, and the right lead, and we have demonstrated explicitly that the
joint effect of dressing $pf$ Green functions as well as exchange
vertices with voltage induced particle-hole excitations works to cut
off these logarithms by $\Gs=\pi\glr^{2}V$.  Under certain conditions,
the T-matrix can be identified with the spectral function on the
quantum dot, which can be measured directly by tunneling into the
dot\cite{Franceschi02}.

To reveal the physical significance of this rate, we have calculated
the dynamical transverse spin susceptibility in the presence of a
finite bias-voltage. This served to demonstrate that $\Gs$ is
indeed the spin-relaxation rate, broadening the resonance pole at
$\w\sim B$ in this correlation function. $\Gs$ arises from the
stirring up of inter-lead particle-hole excitations, and is found to
be proportional, in order $g^2$, to the number of conduction electrons
passing the constriction per unit time (the factor of proportionality
depends, however, on details of the model, such as e.g. anisotropies of
$J$). We therefore interpret the subsequent attenuation of the Kondo
effect as decoherence due to current-induced noise.

Most formulations of perturbative renormalization group in equilibrium
completely neglect the role of decoherence and noise and focus instead
on the flow of coupling constants. This is justified, as the typical
rates are often much smaller than temperature $T$, which serves as the
relevant infrared cutoff. However, since this is not the case in a
nonequilibrium situation, decoherence has to be an essential
ingredient in any formulation of perturbative renormalization group
valid out of equilibrium\cite{Schoeller00,Rosch03a}. We hope that
our perturbative calculation, demonstrating how this happens in
detail, can serve as a starting point for future developments in this
direction.

\begin{acknowledgments}
  
  J. P. acknowledges the hospitality of the {\O}rsted Laboratory at
  the University of Copenhagen, where parts of this work was carried
  out. This work was supported in part by the Center of Functional
  Nanostructures (J.P and P.W.) and the Emmy Noether program (A.R.) of
  the DFG. Additional Funding by the German-Israeli-Foundation is
  gratefully acknowledged.

\end{acknowledgments}

\begin{appendix}

\section{Vertex corrections from crossed rungs}\label{app:crossed}

\begin{center}
\begin{figure}[t]
\includegraphics[width=5cm]{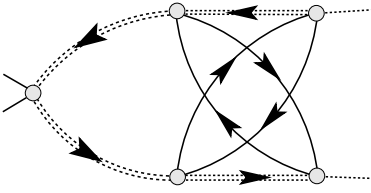}
\caption{\label{fig:vertcross}
  Vertex correction from crossed particle-hole excitations. Such
  contributions are smaller than the ladder-type corrections by a
  factor of $\G/V$ and are therefore neglected.}
\end{figure}
\end{center}

In this Appendix, we argue that crossed diagrams of the form shown in
Fig.~\ref{fig:vertcross} lead only to subleading corrections.  In such
diagrams, for example, the simple frequency dependence observed in
Sec.~\ref{sec:ladder} is no longer valid. We therefore evaluate
explicitly the crossed 4th order correction depicted in
Fig.\ref{fig:vertcross} and compare it to (\ref{eq:vertfour}).

The Feynman rules give the same prefactors in this case, and the
contraction of spins yields
\begin{equation}
(\tau^{i}\tau^{j}\tau^{k}\tau^{m}\tau^{n})_{\al'\al}
{\rm Tr}[\tau^{m}\tau^{j}]
{\rm Tr}[\tau^{i}\tau^{n}]=20\,\tau^{k}_{\al'\al},
\end{equation}
as opposed to $4\,\tau^{k}_{\al'\al}$ obtained in the ladder-type
correction.  The Keldysh contraction may be expressed in terms of the
previously defined tensors $V$ and $B$ as
\begin{eqnarray}
\lefteqn{\hspace*{-10mm}
V{_{\!a'\!b'}}\!\!\!\!\!{^{\!\!c\,d}}\,(\W+\w,\w)
B{_{b''\!a}}\!\!\!\!\!{^{\!\!a' b'''}}\!(\w'-\w)
{\undl {\cal G}}^{b'''a''}\!(\W+\w')}\nonumber\\
&&\times\,
{\undl {\cal G}}^{b''a'''}\!(\w+\w''-\w')
B{_{b'\!a'''}}\!\!\!\!\!\!{^{\!\!\!\!\!a'' b}}\,\,(\w''-\w'),
\end{eqnarray}
and using the identity (\ref{eq:simpleV}), this may be worked out to
give
\begin{eqnarray}
\lefteqn{\hspace*{-8mm}
{^{0\!\!\!\!\!}}_{\times\!}{\tilde \La}{_{ab}}\!\!\!\!\!{^{cd}}^{(4)}
(\W+\w'',\w'')=
\frac{5}{2}\,\delta_{cd}\tau^{3}_{aa}
\!\int\!\frac{d\w'}{2\pi}
\frac{2\Pi^{>}(\w')}{\W+i\G}}\nonumber\\
&&\times\,
2\Pi^{>}(\w''-\w')\GR(\W+\w')\GR(\w'-\w'').\label{eq:crossed}
\end{eqnarray}

This result is reached after eliminating a number of contractions
which vanish under projection, that is, all terms proportional to one
or more factors of $\GL$, and apart from an overall factor of 5,
coming from the different spin sum, this contribution looks very much
like the ladder-type correction (\ref{eq:vertfour}). However, the
second pair of $pf$ Green functions have the structure of an $RR$
product as a function of $\w'$, and when integrated with $\Pi^{>}$
this makes (\ref{eq:crossed}) smaller than (\ref{eq:vertfour}) by a
factor $\G/V$.

Notice that, in contrast to the ladder diagrams, the crossed diagram
in Fig.~\ref{fig:vertcross} involves a loop-integral over ${\cal G}
{\cal G}{\cal G}\Pi$, which does not warrant the omission of $RR$ and
$AA$ terms leading to (\ref{eq:simpleV}). However, keeping all terms
in the $V$-tensor, a rather lengthy contraction leads to a result
which differs somewhat from (\ref{eq:crossed}), but nevertheless
remains smaller than (\ref{eq:vertfour}) by a factor $\G/V$. We expect
higher order corrections to work in the same way, which renders the
ladder-type corrections dominant in $\G/V$ to all orders in the
coupling.

\section{Contractions for ${\rm T}^{R(3)}$}\label{app:contract}

In this Appendix, we work out the contraction of Keldysh indices in Eq.
(\ref{eq:cese3}). There is a total of 9 different non-zero contraction
of $ce$ Keldysh-indices, each of which involve renormalization of
either zero, one, two or all three vertices. This gives rise to a
total of $2^{3}=8$ different types of $pf$ traces, which we need to
work out. If the two $ce$ Keldysh indices are different, a vertex
contributes with a factor of $\delta_{ab}$ rather than $L_{ab}$. Thus
a term with all three vertices renormalized contributes with ${\rm
  Tr}[\Lpf\Gpf\Lpf\Gpf\Lpf\Gpf]$, whereas a term with no vertices
renormalized contributes ${\rm Tr}[\delta\Gpf\delta\Gpf\delta\Gpf]$.
Our strategy will be to perform the contraction and the loop-integral
over $\w$ without including the $\la$-dependent part of the vertex.
After this has been done, it will be a simple matter to include the
additional effects of ${^{0\!}}\Lpf$, by going through very similar
steps once more.

We begin by listing a few useful facts about the relevant matrix
products:
\begin{equation}
{^{0\!}}\Lpf\Gpf=\left(\!\begin{array}{cc}
\GR & 2\GL \\ \phi\GR &  2\phi\GL-\GA
\end{array}\!\right),\,\,\,\,
\delta\Gpf=\left(\!\begin{array}{cc}
\GR & 2\GL \\ 0 & \GA
\end{array}\!\right)
\end{equation}
and
\begin{eqnarray}
\lefteqn{\hspace*{-12mm}{\rm Tr}\left[
\left(\begin{array}{cc}
a_1 & b_1 \\ c_1 & d_1
\end{array}\right)
\left(\begin{array}{cc}
a_2 & b_2 \\ c_2 & d_2
\end{array}\right)
\left(\begin{array}{cc}
a_3 & b_3 \\ c_3 & d_3
\end{array}\right)
\right]=}\nonumber\\
&& \nonumber\\\vspace*{-3mm}
&&
a_1 a_2 a_3 + b_1 c_2 a_3 + a_1 b_2 c_3 + b_1 d_2 c_3 + \nonumber\\
&&
c_1 a_2 b_3 + d_1 c_2 b_3 + c_1 b_2 d_3 + d_1 d_2 d_3.
\end{eqnarray}
The {\it lesser} component Green function takes the form
$\GL=n_{\la}(\GA-\GR)$, and neglecting their slow frequency dependence
we may consider the $pf$ distribution functions as constant
prefactors. This allows us to expand all terms in products of three
Green functions which are either retarded or advanced, and to use
rules like $\GR_1 \GR_2 \GR_3=\GA_1 \GA_2 \GA_3=0$, implied by the
subsequent loop integration which can now be performed by closing in
the half-plane with no poles. Notice that including the frequency
dependence in either factors of $n_{\la}$ or $\undl{\Pi}^{ab}$, coming
from either propagators or vertices, would render such loop-integrals
non-zero. Nevertheless, these contributions will be smaller than the
terms which we retain by a factor $\G/V$ and can therefore be
neglected. Furthermore, the projection allows us to neglect terms
which are proportional to $\GL\GL$ or $\GL\GL\GL$.

With these few rules at hand one may work out the following catalog:
\begin{eqnarray}
\lefteqn{
{\rm Tr}
\left[({^{0\!}}\Lpf\Gpf)_1
({^{0\!}}\Lpf\Gpf)_2({^{0\!}}\Lpf\Gpf)_3\right]=}\\
&&\hspace{30mm}4 n_{\la}\left\{
\phi_1(\GR_1\GR_2\GA_3-\GR_1\GA_2\GA_3)\right.\nonumber\\
&&\hspace*{36mm}
+\phi_2(\GA_1\GR_2\GR_3-\GA_1\GR_2\GA_3)\nonumber\\
&&\left.\hspace*{36mm}
+\phi_3(\GR_1\GA_2\GR_3-\GA_1\GA_2\GR_3)\right\},\nonumber\\
\lefteqn{
{\rm Tr}\left[({^{0\!}}\Lpf\Gpf)_1({^{0\!}}\Lpf\Gpf)_2
(\delta\Gpf)_3\right]=
}\nonumber\\\
&&\hspace{30mm}4 n_{\la}\left\{
\phi_1\GR_1\GA_2\GA_3+\phi_2\GA_1\GR_2\GR_3\right\},\nonumber\\
\lefteqn{
{\rm Tr}\left[({^{0\!}}\Lpf\Gpf)_1(\delta\Gpf)_2
({^{0\!}}\Lpf\Gpf)_3\right]=
}\nonumber\\\
&&\hspace{30mm}4 n_{\la}\left\{
\phi_1\GR_1\GR_2\GA_3+\phi_3\GA_1\GA_2\GR_3\right\},\nonumber\\
\lefteqn{
{\rm Tr}\left[(\delta\Gpf)_1({^{0\!}}\Lpf\Gpf)_2
({^{0\!}}\Lpf\Gpf)_3\right]=
}\nonumber\\\
&&\hspace{30mm}4 n_{\la}\left\{
\phi_2\GA_1\GR_2\GA_3+\phi_3\GR_1\GA_2\GR_3\right\}\nonumber.
\end{eqnarray}
The remaining four possibilities all vanish, and we are left with
contributions from terms with either two or three vertices
renormalized. Working out the loop-integral over $\w$, we get e.g.
\begin{eqnarray}
\int\!\frac{d\w}{2\pi}\,\phi_1\GR_1\GR_2\GA_3&=&\int\!\frac{d\w}{2\pi}
\phi(-\e)\GR(\w)\nonumber\\
&&\hspace*{9mm}\times\GR(\w+\e')\GA(\w+\e)\nonumber\\
&=&i\GA_{\G_2}(\e)\GA_{\G}(\e-\e'),
\end{eqnarray}
where we have introduced the notation $\GA_{\G_2}(\e)=(\e-i\Gs)^{-1}$,
and $\GA_{\G}(\e)=(\e-i\G)^{-1}$ for the {\it double-broadened} $pf$
Green functions.  We see that the vertex corrections serve to replace
$\G$ by $\Gs$ in products of certain internal Green functions, and
working out all the integrals, we obtain the following list for the
Peierls-channel:
\begin{eqnarray}
\int\!\frac{d\w}{2\pi}\,\phi_1\GR_1\GR_2\GA_3&=&
 i\GA_{\G_2}(\e)\GA_{\G}(\e-\e'),
\nonumber\\
\int\!\frac{d\w}{2\pi}\,\phi_1\GR_1\GA_2\GA_3&=&
-i\GA_{\G_2}(\e)\GA_{\G}(\e'),
\nonumber\\
\int\!\frac{d\w}{2\pi}\,\phi_2\GA_1\GR_2\GR_3&=&
 i\GR_{\G}(\e)\GR_{\G_2}(\e'),
\nonumber\\
\int\!\frac{d\w}{2\pi}\,\phi_2\GA_1\GR_2\GA_3&=&
 i\GR_{\G_2}(\e')\GA_{\G}(\e-\e'),
\nonumber\\
\int\!\frac{d\w}{2\pi}\,\phi_3\GR_1\GA_2\GR_3&=&
-i\GA_{\G}(\e')\GR_{\G_2}(\e-\e'),
\nonumber\\
\int\!\frac{d\w}{2\pi}\,\phi_3\GA_1\GA_2\GR_3&=&
-i\GR_{\G}(\e)\GR_{\G_2}(\e-\e').
\end{eqnarray}
As may be seen from Eq.~(\ref{eq:cese3}), the corresponding products
for the Cooper-channel can be obtained from these by the shift of
variables $\e\to-\e'$, and $\e'\to-\e$. Using the fact that
$\GR(-\e)=-\GA(\e)$, one readily obtains the following list, to be
used for the Cooper-channel:
\begin{eqnarray}
\int\!\frac{d\w}{2\pi}\,\phi_1\GR_1\GR_2\GA_3&=&
-i\GR_{\G_2}(\e')\GA_{\G}(\e-\e'),
\nonumber\\
\int\!\frac{d\w}{2\pi}\,\phi_1\GR_1\GA_2\GA_3&=&
-i\GR_{\G_2}(\e')\GR_{\G}(\e),
\nonumber\\
\int\!\frac{d\w}{2\pi}\,\phi_2\GA_1\GR_2\GR_3&=&
 i\GA_{\G}(\e')\GA_{\G_2}(\e),
\nonumber\\
\int\!\frac{d\w}{2\pi}\,\phi_2\GA_1\GR_2\GA_3&=&
-i\GA_{\G_2}(\e)\GA_{\G}(\e-\e'),
\nonumber\\
\int\!\frac{d\w}{2\pi}\,\phi_3\GR_1\GA_2\GR_3&=&
 i\GR_{\G}(\e)\GR_{\G_2}(\e-\e'),
\nonumber\\
\int\!\frac{d\w}{2\pi}\,\phi_3\GA_1\GA_2\GR_3&=&
i\GA_{\G}(\e')\GR_{\G_2}(\e-\e').
\end{eqnarray}

It is now straightforward to carry out the contraction of $ce$ Keldysh
indices in Eq.(\ref{eq:cese3}), and one finds the combination
\begin{eqnarray}
&&\hspace*{2.9mm}
G^{R}G^{R}{\rm Tr}
\left[({^{0\!}}\Lpf\Gpf)_1({^{0\!}}\Lpf\Gpf)_2
({^{0\!}}\Lpf\Gpf)_3\right]
\nonumber\\
&&+G^{K}G^{R}{\rm Tr}
\left[({^{0\!}}\Lpf\Gpf)_1({^{0\!}}\Lpf\Gpf)_2(\delta\Gpf)_3\right]
\nonumber\\
&&+(G^{R}G^{K}+G^{K}G^{A})
{\rm Tr}
\left[({^{0\!}}\Lpf\Gpf)_1(\delta\Gpf)_2
({^{0\!}}\Lpf\Gpf)_3\right]\nonumber
\end{eqnarray}
for the Peierls, and
\begin{eqnarray}
&&\hspace*{2.9mm}
G^{R}G^{R}{\rm Tr}
\left[({^{0\!}}\Lpf\Gpf)_1({^{0\!}}\Lpf\Gpf)_2
({^{0\!}}\Lpf\Gpf)_3\right]
\nonumber\\
&&+G^{K}G^{R}{\rm Tr}
\left[({^{0\!}}\Lpf\Gpf)_1({^{0\!}}\Lpf\Gpf)_2
(\delta\Gpf)_3\right]
\nonumber\\
&&+(G^{R}G^{K}+G^{K}G^{A})
{\rm Tr}
\left[(\delta\Gpf)_1({^{0\!}}\Lpf\Gpf)_2
({^{0\!}}\Lpf\Gpf)_3\right]\nonumber
\end{eqnarray}
for the Cooper-channel. Together, the two channels contribute the
integral
\begin{eqnarray}
\lefteqn{\hspace*{-1mm}{\rm T}_{\g\g'}^{R}{^{(3)}}(\W)=
\frac{3}{16}n_{\la}J^{3}_{\g\g''\g'''\g'}
\!\int\!\frac{d\e}{2\pi}\!\int\!\frac{d\e'}{2\pi}}
\label{eq:tmatint0}\\
&&\times\left\{G_{\g''}^{K}(\W+\e)G_{\g'''}^{R}(\W+\e')
\GA_{\Gs}(\e)\GA_{\G}(\e')\right.\nonumber\\
&&\hspace*{4mm}
-\left[G_{\g''}^{R}(\W+\e)G_{\g'''}^{K}(\W+\e')\right.\nonumber\\
&&\left.\left.\hspace*{7.5mm}
+G_{\g''}^{K}(\W+\e)G_{\g'''}^{A}(\W+\e')\right]
\GA_{\Gs}(\e)\GA_{\G}(\e-\e')\right\}.\nonumber
\end{eqnarray}

To include the effects of ${^{\la\!}}\Lpf$, one may go through the
same steps and build up a similar catalog of terms. We have to
include all terms with exactly one factor of ${^{\la\!}}\Lpf$, since
terms with two or three factors vanish faster than $\langle
Q\rangle_{\la}$ under projection. To leading order in $\G/V$, there
will still only be contributions with either two or three vertices
renormalized. Whereas ${^{0\!}}\Lpf$ ended up contributing only with
its $21$-entry, $\phi$, this entry is zero in ${^{\la\!}}\Lpf$ and
instead one finds only contributions from its $12$-entry, $\psi$. A
typical contribution from the Peierls-channel now takes the form
\begin{eqnarray}
\int\!\!\frac{d\w}{2\pi}{\rm Tr}\!
\left[\!({^{0\!}}\Lpf\Gpf)_1({^{\la\!}}\Lpf\Gpf)_2
(\delta\Gpf)_3\right]
\!\!\!&=&\!\!2n_{\la}\!\!\int\!\!\frac{d\w}{2\pi}
\phi_1\GR_1\psi_{2}\GA_2\GA_3\nonumber\\
&=&\!\!4n_{\la}\frac{i\Gv}{\e'-i\Gs}\GA_{\G}(\e')\GA_{\Gs}(\e),
\nonumber
\end{eqnarray}
and a term like this eventually adds up with a similar term from ${\rm
  Tr}[({^{0\!}}\Lpf\Gpf)_1({^{0\!}}\Lpf\Gpf)_2(\delta\Gpf)_3]$, having
a $1$ in place of the factor of $i\Gv/(\e'-i\Gs)$, to contribute $4
n_{\la}\GA_{\Gs}(\e')\GA_{\Gs}(\e)$.  Working out the full
contribution, from both the Peierls, and the Cooper-channel, one finds
that all surviving terms combine in similar ways, and the total effect
of including ${^{\la\!}}\Lpf$ is therefore simply to replace $\G$ by
$\Gs$ in (\ref{eq:tmatint0}). This finally leads to the integral
(\ref{eq:imcese3}) quoted in the main text.

\section{Cutting off x-ray edge singularities in the anisotropic Kondo
  model}\label{app:xray}

In this Appendix, we investigate the anisotropic Kondo model in the
case of a {\em vanishing} spin-flip coupling ${^{\bot\!\!}}J=0$ and
{\em finite} $^{z\!\!}J$. In this limit, certain equilibrium
correlation functions are singular at the Fermi energy, they display
the so-called X-ray edge singularities whenever the spin is flipped.
In the following, we investigate how these singularities are modified
in the case of a finite voltage.

Even for ${^{\bot\!\!}}J=0$ a finite current is flowing through the
system as $^{z\!\!}J_{LR}\neq 0$ and we therefore expect that the
associated noise will cut off all singularities. Fortunately, a very
similar problem has been solved exactly by Ng\cite{Ng96} (see also
Refs.~\onlinecite{Combescot00} and \onlinecite{Muzykantskii03}), who
considered the effects of suddenly switching on the tunneling between
two (non-interacting) leads.

We will show, that our problem (for ${^{\bot\!\!}}J=0$) can be mapped
exactly on the one solved by Ng. The fact that this is possible is not
obvious as he considered a situation where for times $t<t_{i}$ no
current is flowing, whereas in our case the same current passes the
dot before and after the spin-flip .

Ng considered the Hamiltonian\cite{Ng96}
\begin{equation} \label{hNg}
H_{\rm x}=H_{0}(V)+ \!\!\!\!\!\sum_{\g,\g',{\bf k},{\bf k}',\s,\s'}
\!\!\!\!\!V_{\g'\g}\,c^{\dagger}_{\g'{\bf k}'}c_{\g{\bf k}}
\theta(t_{f}-t)\theta(t-t_{i}),
\end{equation}
where $H_0(V)=\sum_{\g,{\bf k},\s}(\e_{{\bf k}}-\mu_{\g})
c^{\dagger}_{\g{\bf k}\s}c_{\g{\bf k}\s}$ describes the two leads with
the bias voltage $V=\mu_L-\mu_R$. The tunneling between the left and
the right lead (and a potential scattering) is switched on for times
between $t_{i}$ and $t_{f}$. This generalization of the usual X-ray
edge problem to two different Fermi seas was solved by Ng\cite{Ng96},
using a generalization of the method devised by Nozi\`{e}res and De
Dominicis for the problem with only a single Fermi sea. He finds that
the relevant spectral function exhibits power-law singularities near
each of the two Fermi energies in the left and right leads, which are,
however, cut off by a voltage induced broadening given in terms of
complex phase-shifts (see Ref.~\onlinecite{Ng96} for details)
$\delta_{L/R}$, by
\begin{equation}\label{gammaNg}
\G_{\!{\rm x}}=\frac{V}{2\pi}\IM[\delta_{L}-\delta_{R}].
\end{equation}

For ${^{\bot\!}}g=0$ and $B=0$, the Kondo Hamiltonian
(\ref{eq:hamilton}) reduces to two separate potential scattering
problems for conduction electrons of spin up and down, respectively,
\begin{equation}
H=H_0(V)+\!\!\!\!\!\sum_{\g,\g',{\bf k},{\bf k}',\s,\s'}\!\!\!\!\!\!
\left({^{z\!\!}}J_{\g'\g}\,S^{z}/2\right)c^{\dagger}_{\g'{\bf k}'\s'}
\tau^{3}_{\s'\s}c_{\g{\bf k}\s}.\label{eq:ham2}
\end{equation}
and we want to study the effect of a single spin-flip, i.e.
correlation functions like $\langle S^-(t_f) S^+(t_i)\rangle$ or
$\langle [c^\dagger_{\alpha' \uparrow}(t_f) S^-(t_f)] \, [S^+(t_i)
c_{\alpha \uparrow}(t_i)] \rangle$ (which is related to the T-matrix).
For these correlation functions, the spin points down for $t<t_i$,
i.e. $S_{z}=-1/2$ and $H(t<t_i)=H_i=H_0(V)-\frac{1}{4}\sum_{\g,\g',
  {\bf k},{\bf k}',\s,\s'} {^{z\!\!}}J_{\g'\g} c^{\dagger}_{\g'{\bf
    k}'\s'} \tau^{3}_{\s'\s}c_{\g{\bf k}\s}$.  To map
Eq.~(\ref{eq:ham2}) onto Eq.~(\ref{hNg}) we note that $S_{z}=1/2$ for
$t_{i}<t<t_{f}$ and therefore
\begin{equation}
H=H_{i}+
\!\!\!\!\!\sum_{\g,\g',{\bf k},{\bf k}',\s,\s'}\!\!\!\!\!\!
({^{z\!\!}}J_{\g'\g}/2)c^{\dagger}_{\g'{\bf k}'\s'}
\tau^{3}_{\s'\s}c_{\g{\bf k}\s}\theta(t_{f}-t)\theta(t-t_{i}).
\label{kondoNg}
\end{equation}
$H_i$ can easily be diagonalized in terms of scattering states.
Scattering states coming from the left (right) lead are occupied
according to the left (right) chemical potential and therefore,
(\ref{kondoNg}) takes the form (\ref{hNg}) when rewritten in terms of
those scattering states.

To determine the scattering states of $H_i$, we represent for
convenience the two semi-infinite leads by infinite chiral wires of
right-movers. In this representation, the scattering wave-functions
$\Phi^{\g'\g}_{k\sigma}(x)$ describe the  amplitude of plane waves 
coming from lead $\g$ 
\begin{equation}
\Phi^{\g'\!\g}_{k\sigma}(x)=\left[\theta(-x)\delta_{\g'\g}
+\theta(x)S_{\g'\!\g}\right] e^{i k x},
\end{equation}
where $x<0$ ($x>0$) refers to incoming (outgoing) waves in lead $\g'$.
The scattering matrix $S_{\g'\!\g}$ is determined from the
Schr\"odinger equation
\begin{equation}
\left[-i v_{F}\partial_{x}\delta_{\g'\g''} - 
\frac{\sigma}{4}{^{z\!\!}}J_{\g'\g''}\delta(x)\right]
\Phi_{k\sigma}^{\g''\!\g}(x)=
\varepsilon\,\Phi_{k\sigma}^{\g'\!\g}(x),
\end{equation}
and regularizing the delta-function by using $\theta(0)=1/2$ we obtain
\begin{eqnarray}
S_{LL}&=&
\frac{1-(\gz_{LR}^2-\gz_{LL}\gz_{RR})/64+i\sigma (\gz_{LL}-\gz_{RR})/8}
     {1+(\gz_{LR}^2-\gz_{LL}\gz_{RR})/64-i\sigma(\gz_{LL}+\gz_{RR})/8},
\nonumber\\
S_{LR}&=&
\frac{\gz_{LR} 2 i \sigma /8}
     {1+(\gz_{LR}^2-\gz_{LL}\gz_{RR})/64-i\sigma(\gz_{LL}+\gz_{RR})/8},
\nonumber\\
\end{eqnarray}
with $\gz_{\g\g'}=N(0)\,{^{z\!\!}}J_{\g\g'} $ and $N(0)=1/v_F$.

Rewriting Eq.~(\ref{kondoNg}) in terms of these scattering states,
  we can read off the potential in Eq.~(\ref{hNg})
\begin{eqnarray}
V_{\g'\g}=\frac{1}{2}
\sum_{\beta\beta'}\,{^{z\!\!}}J_{\beta'\beta}\,
[\Phi_{\s}^{\beta'\g'}(0)]^{\ast}\,
 \Phi_{\s}^{\beta\g}(0).
\end{eqnarray}
Using this formula and the results by Ng\cite{Ng96}, one can easily
work out the relevant correlation functions when taking into account
that the spin-up and spin-down problems separate. The corresponding
correlation functions are therefore multiplied in the time-domain and
convoluted as a function of frequency. We will not display the rather
lengthy formulas, but only note that all divergences close to the two
Fermi levels are cut off by the appropriate relaxation
rates~(\ref{gammaNg}) [the rates for spin-up and spin-down add as
$e^{-\Gamma_\uparrow t} e^{-\Gamma_\downarrow t}
=e^{-(\Gamma_\uparrow+\Gamma_\downarrow) t}$].

To make contact with our perturbative results, we will now consider
the case of small $^{z\!\!}J$. In this limit $V_{\g'\g}\approx\,
{^{z\!\!}}J_{\beta'\beta}/2$. Inserting this into Eqs. (11d) and (11f)
of Ref.~\onlinecite{Ng96}, determining the complex phase-shifts
$\delta_{L/R}$, expanding the result to leading order in $V_{\g'\g}$
and adding spin-up and spin-down contributions, we find
\begin{equation}
\G_{\!{\rm x}}=\frac{\pi}{2}V|{^{z\!}}g_{LR}|^{2},
\end{equation}
which coincides with our $\Gt=1/T_2$ in Eq.~(\ref{gamma2Aniso}), in
the limit of ${^{\bot\!}}g\to 0$. Note that the first logarithm in
Eq.~(\ref{anisotropicT}) arises from a diagram with $S_z$ at an
external vertex. Therefore the corresponding correlator is not of the
X-ray edge form discussed in this Appendix.

\end{appendix}



\begin{thebibliography}{99}
  
\bibitem{Hewson93} A. C. Hewson, {\it The Kondo Problem to Heavy
    Fermions}, Cambridge University Press (1993).
  
\bibitem{Experiments}D. Goldhaber-Gordon, Hadas Shtrikman, D. Mahalu,
  David Abusch-Magder, U. Meirav and M. A. Kastner, Nature {\bf 391},
  156 (1998); S. M. Cronenwett, T. H. Oosterkamp and L. P.
  Kouwenhoven, Science {\bf 281}, 540 (1998); W. G. van der Wiel, S.
  De Franceschi, T. Fujisawa, J. M. Elzerman, S. Tarucha and L. P.
  Kouwenhoven, Science {\bf 289}, 2105 (2000); J. Nyg{\aa}rd, D. H.
  Cobden and P. E. Lindelof, Nature {\bf 408}, 342 (2000).

\bibitem{Korringa50}J. Korringa, Physica {\bf 16}, 601 (1950).
  
\bibitem{Wang68}Y.-L. Wang and D. J. Scalapino, Phys. Rev. {\bf 117},
  734 (1968).
  
\bibitem{Schoeller00} H. Schoeller and J. K{\"o}nig, Phys. Rev. Lett.
  {\bf 84}, 3686 (2000); M.~Keil, ph.d. thesis, Aachen (2002); H.
  Schoeller, Lect.Notes Phys. {\bf 544}, 137 (2000).
  
\bibitem{Rosch03a}A. Rosch, J. Paaske, J. Kroha and P. W\"olfle, Phys.
  Rev. Lett. {\bf 90}, 076804 (2003)
  
\bibitem{Wolf69}E. L. Wolf and D. L. Losee, Phys. Rev. Lett. {\bf 23},
  1457 (1969).
  
\bibitem{Appelbaum66}J. Appelbaum, Phys. Rev. Lett. {\bf 17}, 91
  (1966); Phys. Rev. {\bf 154}, 633 (1967).

\bibitem{Appelbaum72}J. Appelbaum and L. Y. Shen Phys. Rev. B {\bf 5},
  544 (1972).
  
\bibitem{Bermon78}S. Bermon, D. E. Paraskevopoulos and P. M. Tedrow,
  Phys. Rev. B {\bf 17}, 2110 (1978).
  
\bibitem{Meir93}Y. Meir, N. S. Wingreen and P. A. Lee, Phys. Rev.
  Lett. {\bf 70}, 2601 (1993); N. S. Wingreen and Y. Meir, Phys. Rev.
  B {\bf 49}, 11040 (1994).

\bibitem{Rosch01}A. Rosch, J. Kroha and P. W\"{o}lfle, Phys. Rev.
  Lett. {\bf 87}, 156802 (2001).

\bibitem{Kaminski99}A. Kaminski, Yu. V. Nazarov and L. I. Glazman,
  Phys. Rev. Lett {\bf 83}, 384 (1999); Phys. Rev. B {\bf 62}, 8154
  (2000).

\bibitem{Coleman01} P. Coleman, C. Hooley and O. Parcollet, Phys. Rev.
  Lett. {\bf 86}, 4088 (2001); P. Coleman and W. Mao,
  cond-mat/0203001.
  
\bibitem{Kiselev02} M. N. Kiselev, K. Kikoin, and L. W. Molenkamp
  Phys. Rev. B {\bf 68}, 155323 (2003); JETP Letters {\bf 77}, 366
  (2003).
  
\bibitem{Mao03}W. Mao, P. Coleman, C. Hooley and D. Langreth, Phys.
  Rev. Lett. 91, 207203 (2003).
  
\bibitem{Shnirman03}A. Shnirman and Y. Makhlin, Phys. Rev. Lett. 91,
  207204 (2003).
  
\bibitem{Paaske03a}J. Paaske, A. Rosch, and P. W\"{o}lfle,
  cond-mat/0307365.
  
\bibitem{Rammer86}J. Rammer and H. Smith, Rev. Mod. Phys. {\bf 58},
  323 (1986).
  
\bibitem{Walker68}M. B. Walker, Phys. Rev. {\bf 176}, 432 (1968);
  Phys. Rev. B {\bf 1}, 3690 (1970).

\bibitem{Woelfle71} W. G\"otze and P. W\"olfle, JLTP {\bf 5}, 575
  (1971).
  
\bibitem{Langreth72}D. Langreth and J. Wilkins, Phys. Rev. B {\bf 6},
  3189 (1972).
  
\bibitem{Langreth76}D. C. Langreth in {\it Linear and Nonlinear
    Electron Transport in Solids}, eds. J. T. Devreese and E. Van
  Doren (Plenum, New York, 1976); H. Haug and A. -P. Jauho, {\it
    Quantum Kinetics in Transport and Optics of Semiconductors},
  Springer-Verlag (Berlin, 1996).

\bibitem{Rosch03b} A. Rosch, T. A. Costi, J. Paaske, P. W\"olfle, Phys.
  Rev. B {\bf 68}, 014430 (2003).
  
\bibitem{Franceschi02} S. De Franceschi, R. Hanson, W. G. van der
  Wiel, J. M. Elzerman, J. J. Wijpkema, T. Fujisawa, S. Tarucha, and
  L. P. Kouwenhoven, Phys. Rev. Lett. {\bf 89}, 156801 (2002).
  
\bibitem{Ng96}Tai-Kai Ng, Phys. Rev. B {\bf 51}, R2009 (1995); {\bf
    54}, 5814 (1996).
  
\bibitem{Combescot00}M. Combescot and B. Roulet, Phys. Rev. B {\bf
    61}, 7609 (2000).

\bibitem{Muzykantskii03}B. Muzykantskii, N. d'Ambrumenil and B.
  Braunecker, cond-mat/0304583.
  
\end{thebibliography}
\end{document}